\documentclass{aa}  

\usepackage{graphicx}
\usepackage{placeins} % for \Floatbarrier
\usepackage{txfonts}
\usepackage{hyperref}
\usepackage{mathrsfs} % \mathsrc{}
\usepackage{soul}
\usepackage{siunitx}

\usepackage{color} % gestion de différentes couleurs
\usepackage{amssymb}

\newcommand{\Msol}{M_{\odot}}
\newcommand{\vn}{\vec{n}}

\newcommand{\vR}{\vec{R}}
\newcommand{\vr}{\vec{r}}
\newcommand{\vB}{\vec{B}}
\newcommand{\vV}{\vec{V}}

\newcommand{\vP}{\vec{P}}
\newcommand{\vp}{\vec{p}}
\newcommand{\nodot}{\vec{n}_\odot}

\newcommand{\pderiv}[2]{\frac{\partial #1}{\partial #2}} % write the partial derivative as d#1/d#2

\newcommand{\voisin}{\cite{voisin_improved_2020}\ }
\newcommand{\voisinp}{\citep{voisin_improved_2020}\ }

\begin{document} 

   %\title{Timing PSR J0337+1715: the strong equivalence principle with red noise or a Planet }
    \title{Explanation of the exceptionally strong timing noise of PSR J0337+1715 by a circum-ternary planet and consequences for gravity tests}
%   \subtitle{I. Overviewing the $\kappa$-mechanism}

\titlerunning{J0337's extreme timing noise explained by a circum-ternary planet}

   \author{G. Voisin\inst{1}\fnmsep\thanks{Email: guillaume.voisin@obspm.fr}
          \and
          I. Cognard\inst{2}\fnmsep\inst{3} %\thanks{Just to show the usage of the elements in the author field}
          \and 
          M. Saillenfest\inst{4}
          \and 
          T. M. Tauris\inst{5}\fnmsep\inst{6}
          \and 
          N. Wex\inst{6}
          \and
          L. Guillemot\inst{2}\fnmsep\inst{3}
          \and
           G. Theureau\inst{2}\fnmsep\inst{3}\fnmsep\inst{1}
           \and
         P. C. C. Freire\inst{6}
         \and
         M. Kramer\inst{6}
         }

\institute{Laboratoire Univers et Théorie, Observatoire de Paris, Université PSL, Université de Paris, CNRS, F-92190 Meudon, France\\
        \and
        Laboratoire de Physique et Chimie de l'Environnement et de l'Espace, Universit\'e d'Orl\'eans / CNRS, 45071 Orl\'eans Cedex 02, France\\
        \and
        Observatoire Radioastronomique de Nan\c{c}ay, Observatoire de Paris, Universit\'e PSL, Université d'Orl\'eans, CNRS, 18330 Nan\c{c}ay, France \\
        \and
            IMCCE, Observatoire de Paris, Université PSL, CNRS, Sorbonne Université, Université de Lille, 75014 Paris, France\\
        \and
            Dept. of Materials and Production, Aalborg University, DK-9220 Aalborg Øst, Denmark  
        \and
           Max-Planck-Institut f\"{u}r Radioastronomie, Auf dem H\"{u}gel 69, D-53121 Bonn, Germany\\
        %\and
        %    LESIA, Observatoire de Paris, Universit\'e PSL, CNRS, Sorbonne Universit\'e, Universit\'e de Paris, 5 place Jules Janssen, 92195 Meudon, France
             %\thanks{The university of heaven temporarily does not
              %       accept e-mails}
    }
   
   \date{Received 3 September 2024; accepted 12 November 2024}

% \abstract{}{}{}{}{} 
% 5 {} token are mandatory
 
  \abstract
   {Timing of pulsar PSR J0337+1715 provides a unique opportunity to test the strong equivalence principle (SEP) with a strongly self-gravitating object. This is due to its unique situation in a triple stellar system with two white dwarfs.}
    {Our previous study suggested the presence of a strong low-frequency residual signal in the timing data, and we set out to model this signal on a longer dataset in order to determine its nature and improve accuracy.}
 %methods
    {We considered three models: chromatic red noise, achromatic red noise, and a small planet in a hierarchical orbit with the triple stellar system. These models were implemented in our numerical timing model. We performed Bayesian inference of posterior distributions. Best fits were compared using information-theoretic criteria. }
    {We rule out chromatic red noise from dispersion-measure variations. Achromatic red noise or a planet in Keplerian orbit provide the best fits. If the residual signal is red noise, then it appears exceptionally strong.  
   	When assuming the presence of a planet, we obtained a marginal detection of mutual interactions that allowed us to constrain its mass to $\sim 0.5 M_{\rm Moon}$ as well as its inclination. The latter is intriguingly coincident with a Kozai resonance. We show that a longer observation span will ultimately lead to a clear signature of the planet model due to its mutual interactions with the triple system.
    We produce new limits on SEP violation: $|\Delta| < 1.5\cdot 10^{-6}$ or $|\Delta| < 2.3\cdot 10^{-6}$ at a 95\% confidence level under the planet or red-noise hypothesis, respectively. This model dependence emphasises the need for additional data and model selection. As a by-product, we estimated a rather low supernova kick velocity of $\sim 110-125 \rm km/s$, strengthening the idea that it is a necessary condition for the formation of pulsar triple systems.
     }
   {}


   \maketitle
%
%=========================================================================================

\section{\label{sec:intro}Introduction}

Being extremely accurate clocks \citep{manchester_millisecond_2017}, millisecond pulsars (MSPs) undergo particular scrutiny for any timing irregularity that may be the signature of a wide variety of phenomena, including studies of stellar evolution \citep[e.g.][]{tv23,tkf+17}, the discovery of the first extra-solar planets \citep{wolszczan_planetary_1992}, the ongoing pursuit of a low-frequency gravitational-wave background \citep[e.g.][]{reardon_parkes_2021, alam_nanograv_2021, chen_common-red-signal_2021}, constraining the neutron-star equation of state \citep[e.g.][]{Fonseca_2021, riley_nicer_2021}, or tests of gravity theories \citep[e.g.][]{Freire_Wex_2024,kramer_strong-field_2021}.
One of the most prominent systems for the study of stellar evolution and especially for tests of gravity theories is PSR~J0337+1715 (J0337 hereafter). This pulsar was discovered in the GBT drift-scan survey \citep{boyles_green_2013, lynch_green_2013}, and the triple nature of the system was confirmed by \cite{ransom_millisecond_2014}. The pulsar follows a $\sim 1.6$-day orbit with the inner $\sim 0.2M_\odot$ helium white dwarf, which is detectable at optical wavelengths \citep{kaplan_spectroscopy_2014}. Together they form what we call the inner binary. The centre of mass of the inner binary can be seen as orbiting with the outer $\sim 0.4 M_\odot$ white dwarf in $\sim 327$ days. Interestingly, the orbits have low eccentricities and are nearly co-planar, which has proved fundamental for understanding the evolution of the system \citep{tauris_formation_2014}.

Altogether, the triple system extends over $\sim 1$ AU, which makes it uniquely compact compared to other multiple systems. The first pulsar that was found in a triple system was PSR B1620$-$26 \citep{backer_second_1993, thorsett_triple_1999}. The system is composed of a likely gas giant companion orbiting a pulsar white dwarf binary in a hierarchical orbit \citep{sigurdsson_young_2003, sigurdsson_update_2005}. To our knowledge, there are three other systems suspected to be hierarchical triple systems: ZTF J1406+1222, which is a black-widow pulsar candidate that has recently been suggested to be associated with a distant low-mass stellar companion in a $\sim 12,000$-year orbit \citep{burdge_62-minute_2022} extending over $\sim 600$AU; the redback pulsar PSR 2043+1711, which has a suspected distant stellar companion \citep{donlon_anomalous_2024}; and the pulsar binary PSR J1618$-$3921, which shows anomalous orbital-period variations and a spin-period second derivative compatible with the presence of a third object in the system\citep{grunthal_triple_2024}.  There is yet another system, PSR~J1903+0327 \citep{champion_2008}, that is thought to have formed in a triple system that later became unstable \citep{Freire_2011,Portegies-Zwart_2011,pcp12} and is currently a unique type of binary system. 

The presence of an MSP in such a relatively compact triple stellar system allows for a unique test of the universality of free fall (UFF) involving a strongly self-gravitating object \citep{freire_tests_2012, ransom_millisecond_2014, shao_testing_2016, archibald_universality_2018, voisin_improved_2020, voisin_one_2022}.
We now discuss the reasons for this and why it is important (for a more detailed explanation see \cite{voisin_improved_2020}).

The UFF, extended to self-gravitating objects, is the gravitational weak equivalence principle \citep[GWEP;][]{will_confrontation_2014}.  Together with the local Lorentz and position invariances of gravity, it constitutes the strong equivalence principle (SEP).
% PF: this part is not really necessary: 
% If a gravitational equivalent of Schiff's conjecture applies \citep{will_confrontation_2014},  the GWEP implies the validity of the SEP.
Testing the SEP is of great importance because of all valid theories of gravity, only general relativity (GR) fully embodies it;\footnote{The only other SEP-abiding theory is Nordstr\"om’s scalar theory \citep{2007CQGra..24.1867G, deruelle_nordstroms_2011}; however, this was shown to be inconsistent with several experimental results on light deflection.} all others predict some type of SEP violation. 
This means that a violation of the SEP is perhaps the most promising avenue for finding new physics beyond GR. This also means that searching for SEP violation will either rule out GR (in case of a detection) or alternative theories of gravity (in case of a non-detection).

In alternative theories of gravity, such as scalar-tensor theories \citep{damour_nonperturbative_1993, damour_tensor-scalar_1996}, violation of GWEP takes the form of an effective body-dependent gravitational constant $G_{ab} = G_{ba} = G(1+\Delta_{ab})$, where $G$ is the Newtonian gravitational constant and $\Delta_{ab}$ is a violation parameter, the definition of which is theory dependent. In weakly self-gravitating bodies this reduces to the usual discrepancy between inertial and gravitational masses. Assuming WEP applies, the ratio of the two masses is then proportional to the fractional gravitational binding energy, where the proportionality coefficient is the so-called Nordtvedt parameter \citep{nordtvedt_testing_1968}. One can see that in a binary system orbiting according to Newton's equations of motion, substituting $G$ for $G_{ab}$ only amounts to re-scaling masses, and with those being unknown, it does not allow for any constraint on the SEP (which is no longer true with post-Newtonian equations of motion). In a triple system, however, there is no such degeneracy. 
That is why the WEP has been tested using two masses of various compositions falling in the gravitational field of a third one, as in the case of the MICROSCOPE experiment, which had two masses in Earth orbit for one year \citep{Touboul_2022}.

The GWEP was first tested with the Lunar Laser Ranging (LLR) experiment, where the triple system of self-gravitating objects is the Earth-Moon-Sun ($\oplus\, \textnormal{-}\circ\textnormal{-}\, \odot$) system. The extremely accurate determination of the Moon's distance provided by LLR resulted in a tight limit on the Nordtvedt effect \citep{nordtvedt_testing_1968} and a resulting limit of $ \Delta_{\rm \oplus \odot} - \Delta_{\circ \odot} = (3.0\pm5.0)\times 10^{-14}$ \citep{hofmann_relativistic_2018}. However accurate, this result 
provides a weak limit to the SEP (in particular the Nordtvedt parameter) because of the small binding energies of the Earth and Moon. Furthermore, such weak-field limits do not constrain the strong-field regime because of additional theory-dependent effects that may arise, such as spontaneous scalarisation \citep{damour_nonperturbative_1993, damour_tensor-scalar_1996}.

For this reason, experiments with pulsars as the compact object are important. Prior to the discovery of J0337, the best that could be done was the Damour-Sch\"affer test \citep{damour_new_1991}, where the triple system is effectively a near-circular pulsar - white dwarf binary evolving in the gravitational potential of the Galaxy. A violation of GWEP would polarise the orbital eccentricity of the binary towards the Galactic centre.
% PF: This first result (from Gonzalez et al. 2011) was a bit compromised, because it uses a pulsar it should not use, which has an omega-dot rate that is slower than the orbital time around the Galaxy. Nonetheless, these type of tests were then superseded by the work of Zhu et al. 2019, which uses a much more sound method.
% By statistically combining a set of such binaries, \cite{gonzalez_high-precision_2011} obtained $|\Delta_{\rm pG} - \Delta_{\rm cG}| < 4.6\times 10^{-3}$ (95\% C.L.), where $\rm c$ is the companion, and $\rm G$ the Galaxy. Later, 
% PF: So we can just mention the Zhu et al. result, and forget about the previous.
Using one of the most accurately timed pulsars, PSR J1713+0747, \cite{Zhu_testing_2019}
obtained $|\Delta_{\rm pG} - \Delta_{\rm cG}| < 2 \times 10^{-3}$ at a 95\% confidence level (95\% C.L.) from the lack of variation of the orbital eccentricity of the system. This type of test has one main limitation: the weak gravitational acceleration of the Galaxy.
After the discovery of J0337, it became clear that the system is an extremely sensitive probe of GWEP violations given the (locally) much stronger field of the outer white dwarf star compared to the field of the Galaxy.

As was first shown in \cite{ransom_millisecond_2014} and \citet{archibald_universality_2018}, the aforementioned compactness of the J0337 system makes it necessary to numerically integrate the three-body equations of motion at first post-Newtonian order to model the timing data. This represents a large computational cost compared to the usual pulsar-timing modelling, which relies on analytical expressions for binary systems. To perform these numerical integrations, \cite{voisin_improved_2020} independently developed a numerical timing model (\texttt{Nutimo}) \voisinp that supports the strong-field generalisations of the parametrised post-Newtonian equations of motion (for instance \cite{damour_strong-field_1992} and Appendix A of \voisin), which is necessary for a test of alternative theories of gravity, particularly the UFF. J0337's orbital motion is weakly relativistic, and as such, it is not very sensitive to post-Newtonian effects \citep{damour_strong-field_1992}, which are the key to test GR in relativistic pulsar binaries \citep{kramer_strong-field_2021}; this is not a problem because any violation of the GWEP would appear at the Newtonian level in the equations of motion.

Using independent datasets and independent analysis, \cite{archibald_universality_2018} and \voisin\, showed that this system constrains deviations from GWEP to $|\Delta| < 2.6\cdot 10^{-6}$ and $|\Delta| < 2.1\cdot 10^{-6}$, respectively, at a 95\% C.L., where $\Delta$ denotes a violation of GWEP between the pulsar and any of the companions. In both cases, these results represent an improvement of nearly three orders of magnitude compared to the best previously available test of GWEP with a neutron star, using the above-mentioned PSR 1713+0747 \citep{zhu_testing_2015}. This demonstrates how relevant J0337 is for this kind of work. A notable difference between the two approaches is the fact that the limit of \cite{archibald_universality_2018} is mostly based on limits on systematics, while the limit of \voisin is mostly statistical. The latter was additionally able to constrain the longitudes of ascending nodes of the system. 

\paragraph{Motivation of this work:}

Here, we continue the work of \cite{voisin_improved_2020} with an extended timing dataset of J0337 from the Nan\c{c}ay radiotelescope, now spanning approximately eight years (against five previously). The main motivation for the continued timing of this system is to improve the test of the GWEP provided by J0337.

A second motivation for this paper, which is linked with the first one, is to investigate an unmodelled low-frequency signal in the timing of J0337. This signal was already present in \voisin, but it was interpreted as a possible red-noise process intrinsic to the pulsar emission mechanism \cite[e.g.][]{shannon_assessing_2010, lyne_switched_2010} and left unmodelled. Instead, in \voisin uncertainties were widened in order to accommodate this effect and produce a conservative estimate of the posterior distribution functions, particularly that of the $\Delta$ parameter. However, in the more recent data discussed in this work, its amplitude has been increased to $\sim 4 \rm \mu s$, which is to be compared to the $\sim 2 \rm \mu s$ uncertainty on individual times of arrival. This amplitude is such that it systematically limits the GWEP test with J0337. Thus, we aim to model it and evaluate the nature of its cause.

To do this, we considered two types of models for this low-frequency component: 1) red timing noise (chromatic or achromatic) and 2) the presence of a planet in a wide hierarchical orbit around the triple stellar system. The latter option is motivated by the unusually large amplitude of the residual signal compared to other known red-noise processes in millisecond pulsars (see Sect. \ref{sec:discussion}). Precedents of companions initially associated with strong timing noise include the already mentioned planet around the pulsar binary PSR B1620$-$26 \citep{backer_second_1993, thorsett_triple_1999} as well as the likely distant stellar companion to the isolated pulsar PSR J1024$-$0719 \citep{bassa_millisecond_2016, kaplan_psr_2016}. To these can be added the proposal that the noise of PSR J1939+2134 (PSR B1937+21) may be caused by an asteroid belt \cite{shannon_asteroid_2013}. In any case, the apparent periodicity of the signal is comparable with the eight-year observation span, which will render our conclusions concerning its cause necessarily preliminary.

A third motivation for this paper is obtaining a more detailed study of the evolution of the system. Notably, this is also important for evaluating whether the presence of a planet is possible.

%(PF: The following material could be somewhere else, like the start of section 3, otherwise the introduction is way too long).

Since the discovery of PSR B1257+12 \citep{wolszczan_planetary_1992} and its system of two terrestrial-mass and one moon-mass planets \citep{konacki_masses_2003}, only five more pulsars have been found to host planets \citep[][and references therein]{nitu_search_2022,Vleeschower_2024}, four of which are dense Jupiter-mass `diamond planets' that are likely remnant white dwarf cores \citep[][]{bailes_transformation_2011}. The fifth is the planet in the PSR B1620$-$26 system \citep{backer_second_1993, thorsett_triple_1999, sigurdsson_young_2003}. In addition are the already mentioned candidate companion to PSR J1555$-$2908 \cite{nieder_is_2022} and the possible asteroid belt around PSR J1939+2134 (PSR B1937+21) \cite{shannon_asteroid_2013}. Surveys have put limits on the planet population, particularly \cite[][]{kerr_limits_2015} around young pulsars and  \cite{behrens_nanograv_2020} for millisecond pulsars, both of which detected no planets.

More recently \cite{nitu_search_2022} studied a broader sample of 800 pulsars and found 15 candidates, most of which are considered likely to be quasi-periodic noise of magnetospheric origin. In some of these cases simultaneous pulse-profile variations can be observed, clearly pointing at that explanation. 

These examples show the particular difficulty involved in detecting small planets. However, in the case of J0337, a clearer signature can be expected due to the specific mutual interactions within a four-body system.
%This shows again that one of the key difficulties is to disentangle the planet signal from red noise. Indeed, for relatively wide orbits (a few years) the effect of a low-mass companion is only due to the geometric displacement of the pulsar, the so-called R\o mer delay. This signal is purely sinusoidal (with a first harmonic in case of an eccentric orbit), which can easily be absorbed into red-noise provided the signal is weak enough (due to low mass) and the number of observed orbital revolution is few, which for long wide orbits is usually the case. In the case of J0337, however, the more complex orbital configuration can in principle lift this degeneracy, as we show in section \ref{sec:models}. 

In the following sections of this paper, we present the extended dataset in Sect. \ref{sec:observations}, describe the models in Sect. \ref{sec:models}, and present the posterior inference and fitting results in Sect. \ref{sec:results}. We end with a discussion of the consequences of the two main hypotheses, red noise or planet, in Sect. \ref{sec:discussion}.

%-----------------------------------------------------------------------------------------

\section{Observations \label{sec:observations}}

The pulsar J0337+1715 has been regularly observed since July 2013 every two or three days with the Nançay radio telescope (NRT) using its L-band receiver at a central frequency of 1484 MHz. The NRT is a meridian Kraus design collector equivalent to a 94-meter dish able to conduct $\sim 1$ hour observations on any given source within its declination range each day.
The dual linear polarisation signals are sent to the Nan\c cay Ultimate Pulsar Processing Instrument (NUPPI, \cite{2011AIPC.1357..349D,2013sf2a.conf..327C}), an instrument that is able to coherently dedisperse a total bandwidth of 512 MHz. In this work, we use a dataset of 12474 20-mins integrated times of arrival (ToA) divided in four 128 MHz bands observed between MJD 56492 and MJD 59480 (July 2013 and September 2021). \footnote{Dataset, version of \texttt{Nutimo} used in this work, and results are available on Zenodo at \url{https://doi.org/10.5281/zenodo.13899771}. \label{fn:zenodo} } As in the previous analysis presented in \voisin where details can be found, the times of arrival are estimated using the \texttt{pat} tool from the \texttt{PSRCHIVE} software library \citep{hotan_psrchive_2004}, with the Fourier domain with Markov chain Monte Carlo (FDM) method. We note that two more years of data were added for this analysis. 

At MJD 58631, new cooled pre-amplifiers were installed at the telescope, boosting the sensibility in the upper part of the band ($\sim$ 1550-1730MHz) and slightly changing the overall pulse shape (the exact pulse shape is smoothly frequency dependent). Around MJD 58780, we started a better polarisation calibration observing a slow, strong and polarised pulsar over one hour and the receiver rotating for $\sim$180 degrees \citep{guillemot_improving_2023}. Thus the quality of the data determination in the range MJD 58631 -- 58780 has been altered due to inappropriate calibration. In practice, this leads to significantly larger ToA uncertainties during that $\sim$150 days period, reflecting that small discrepancy between the template and the distorted daily profiles. 
As a result, although we do not expect the timing analysis to be significantly biased (thanks to wider uncertainties), we preferred to conservatively excise the range MJD 58631 -- 58780 from our dataset.

\section{Models \label{sec:models}}
The present analysis is based on the numerical timing model of \voisin. In the remainder of this section, we focus on two types of extension of the model aimed at explaining the residual low-frequency signal (see footnote \ref{fn:zenodo}). The first extension assumes that the signal is due to a generic achromatic red-noise process modelled by a Fourier decomposition with a power-law spectrum, and an optional chromatic component caused by non-linear variations of dispersion-measure (DM) over time. The second extension assumes that the signal is caused by a small planet in a hierarchical orbit with the triple system. Parameters associated with the triple system are defined as in \voisin, but nonetheless reviewed in Sec. \ref{sec:planet} such that this paper is self-contained.

All models include astrometric and pulsar spin parameters that were described in\cite[][]{voisin_simulation_2017, voisin_improved_2020}.  Astrometric parameters are right ascension, declination and distance as well as the associated proper motion. Intrinsic parameters are the spin frequency and its derivative re-scaled to absorb the linear and quadratic effects of the Einstein and Shklovskii delays \citep{voisin_improved_2020}. Except in the dedicated chromatic model (see below), dispersion measure is fitted using two parameters, a constant value and a linear drift.

\subsection{Red noise}

Red noise in pulsar timing can be due to several causes. If it is chromatic, then the most common cause are variations of dispersion measure, which are a propagation effect caused by fluctuations of the column density of electrons in the interstellar medium. Such variations can be disentangled from achromatic signals if the observation is spectrally resolved, which is the case in the present work. Other sources of chromatic noise are scattering and band noise, which we do not consider in this work. Achromatic red noise, on the other hand, is intrinsic to the system \citep[e.g.][]{shannon_assessing_2010}, and has been proposed to be caused by magnetospheric variations \cite{lyne_switched_2010, tsang_timing_2013}, changes in the neutron star interior \citep[e.g.][and references therein]{melatos_pulsar_2014}, or by an asteroid belt around the pulsar \citep[][]{shannon_asteroid_2013}. 

The latter case illustrates the difficulty there can be in disentangling possible orbital motion caused by low-mass objects from other sources of noise. A recent example is the proposed presence of a small planet orbiting the black-widow system hosting PSR J1555$-$2908 \citep{nieder_is_2022} in order to explain an unusually strong red noise signal.

\subsubsection{Achromatic red noise}
In recent years, red noise has been modelled for an increasing number of millisecond pulsars in the context of pulsar timing arrays \citep[e.g.][]{chalumeau_noise_2022, alam_nanograv_2021, goncharov_identifying_2021}.
%in order to improve their sensitivity to gravitational waves. This showed that most millisecond pulsars have undetectable red-noise levels, and provides us with indications concerning its likely properties in millisecond pulsars. 
Achromatic red noise is usually modelled as a truncated Fourier series the coefficients of which follow a Gaussian stochastic process with a power-law power-spectrum density. In order to select the best noise model, the posterior distribution function is marginalised over the deterministic part (the rest of the timing model) common to all models and selection is based on the comparison of their evidence using a Bayes factor \citep[see, e.g.][]{chalumeau_noise_2022}. 

We faced several difficulties in the present work: i) We wanted to compare the red-noise model to other models that, involving a planet, are deterministic. ii) We could not perform analytical marginalisations over all the common parameters due to the numerical nature of the model. iii) Evidence computation is computationally expensive. We note that point iii) is largely a consequence of point ii). On the other hand, given the relatively limited observation span, only a few Fourier components are actually necessary, which makes it possible to describe red noise with a deterministic model.

Thus, we added to the timing formula a truncated Fourier series the amplitudes of which follow a power law. This is equivalent to the average spectrum produced by a stochastic Gaussian process as described above. Formally, it can be written as 
\begin{equation}
\label{eq:RN}
    F(t_a) = \sum_{k=1}^n \frac{A}{k^\gamma} \sin\left(2k\pi\nu t_a + \phi_k \right),
\end{equation}
where $A$ is the amplitude of the Fourier component at the fundamental frequency $\nu$, $\gamma$ is the power-law index, $\phi_k$ are phases at each harmonic $k$, and $t_a$ are the times of arrival in the solar-system-barycentre frame. We note that the difference introduced by using times of arrival instead of times of emission is negligible given the time scale and amplitude of the signal.

We considered models from two to five Fourier components that we call `PLn' where $n\in\{2,3,4,5\}$. There are $3+n$ fitted parameters, which are $\nu, A, \gamma, \phi_{k\in[1,n]}$. We remark that, contrary to the stochastic process approach, we fit for the fundamental frequency $\nu$. Indeed, under the prior assumption that the Fourier spectrum is given by a power-law, the stochastic approach evaluates the probability of the signal being described by a complete Fourier series and therefore the fundamental frequency can be fixed to the inverse of the data span. Equation \eqref{eq:RN} is only a truncated series that therefore has additional freedom. Additionally, assuming the signal results from a stochastic process, here we fit a particular realisation of it that is unlikely to follow an exact power-law spectrum, especially one that is truncated. Fitting for the fundamental frequency may partially compensate for this.

\subsubsection{Chromatic contribution: DMX}
In order to test if the measured signal is due to variations in the interstellar medium, we tested a model combining an achromatic red-noise signal as described above with a variable dispersion measure (DM). The observation span was split in ten evenly distributed intervals each with a different $\rm DM_X$ value of DM, where $\rm X$ runs from one to ten. In this way, each interval covers $\simeq 299$ days, which is much smaller than the residual signal timescale of $\sim 3000$ days and can capture it reasonably well if the signal turns out to be chromatic. On the other hand, this time interval is comparable to the orbital period of the outer binary ($P_{\rm O} \simeq 327$ days), which avoids any correlations by averaging any effect due to the outer-binary motion over the interval. 

\subsection{A planet in a hierarchical orbit with the triple stellar system} \label{sec:planet}
A smooth, slow and quasi-sinusoidal signal is expected from a planet of mass $m_\mathrm{\pi}$ orbiting the triple system in a hierarchical orbit, that is with a period $P_\mathrm{\Pi} \gg P_\mathrm{O}$ with $P_\mathrm{O}$ the orbital period of the outer white dwarf, and a mass sufficiently small to ensure that its gravitational field is negligible compared to the fields within the triple system. In the present case, it is sufficient to consider planets of mass $m_\mathrm{\pi} \ll M_\odot$. The only measurable effect given the amplitude of the signal (a few $\rm \mu s$) is the so-called R\o mer delay induced by the planet, that is the variation of the distance between the pulsar and the observer induced by the presence of the planet. 

One of the key difficulties is to disentangle the planet signal from red noise. Indeed, for relatively wide orbits (a few years) the R\o mer delay of a low-mass companion is purely sinusoidal (with a first harmonic in case of an eccentric orbit), which can easily be absorbed into red-noise provided the signal is weak enough (due to low mass) and the number of observed orbital revolution is few, which for long wide orbits is usually the case. In the case of J0337, however, the more complex orbital configuration can in principle lift this degeneracy, as we show below. 

In practice, we have implemented in \texttt{Nutimo} the possibility to add extra bodies the orbits of which are integrated numerically along with the orbits of the pulsar and the two white dwarfs. As in \voisin, we have validated the accuracy of the integration in order to obtain a numerical timing accuracy at the level of a few nanoseconds. 

\subsubsection{Parametrisation}
We extend here the orbital hierarchical parametrisation of the triple system \citep{voisin_improved_2020, archibald_universality_2018} in order to include a fourth body. In doing so, we explain how this parametrisation is related to Jacobi coordinates.

Let $\left(\vR, \vV\right)_{k\in \rm\{p,i,o,\pi\}}$ be the positions and velocities of the pulsar (p), inner white dwarf (i), outer white dwarf (o), and planet ($\rm \pi$) relative to the inertial reference frame associated with their centre of mass. We call `inner binary' the set $\rm I = \{p,i\}$ with centre of mass $\rm b$, `outer binary' the set $\rm O = \{b, o\}$ with centre of mass ($\rm t$), which is also the centre of mass of the triple system, and we call `planetary binary' the set $\rm \Pi = \{t, \pi\}$, the centre of mass of which is also centre of mass of the whole system. 

One can decompose the position of the pulsar as
\begin{equation}
\label{eq:decomporp}
    \vR_{\rm p} = \vR_{\rm p/b} + \vR_{\rm b/t} + \vR_{\rm t},
\end{equation}
and similarly its velocity $\vV_{\rm p}$. 
In \voisin, only the first two terms were present on the right-hand side. They describe the motion of the pulsar relative to the inner-binary centre of mass, and of the inner binary with respect to the centre of mass of the system. Here, we add a third term that describes the motion of the triple system with respect to the centre of mass of the whole system, which includes a planet. Thanks to the hierarchy of the system each term follows an approximately Keplerian motion (see also Appendix \ref{sec:pertubative}), which makes a parametrisation of positions and velocities in terms of osculating orbital elements relevant. In addition, mutual interactions as well as Shapiro and Einstein delays allow one to constrain the four masses. This leads to the following correspondence between the state vectors and masses on one side, and orbital elements on the other side,

\begin{eqnarray}
\label{eq:orbparam}
\left(\begin{matrix}
\left(\vR, \vV\right)_{\rm p/b}, \\
\left(\vR, \vV\right)_{\rm b/t}, \\
\left(\vR, \vV\right)_{\rm t}, \\
m_{\rm p}, m_{\rm i}, m_{\rm o}, m_{\rm \pi} 
\end{matrix}\right) & \leftrightarrow & \left(\begin{matrix}
a_{\rm p}, e_{\rm I}, \omega_{\rm I}, t_{\rm asc p}, i_{\rm I}, \Omega_{\rm p}, \\
a_{\rm b}, e_{\rm O}, \omega_{\rm O}, t_{\rm asc b}, i_{\rm O}, \Omega_{\rm b}, \\
a_{\rm t}, e_{\rm \Pi}, \omega_{\rm \Pi}, t_{\rm asc t}, i_{\rm \Pi}, \Omega_{\rm t}, \\
 m_{\rm i}/m_{\rm p},  P_{\rm I},  P_{\rm O}, P_{\rm \Pi}
\end{matrix}\right),
\end{eqnarray}
where $a_{ x}, e_{ X}, \omega_{ X}, t_{\mathrm{asc} x}, i_{ X}, \Omega_{ x}$  are the semi-major axis, orbital eccentricity, argument of periastron, time of passage at the ascending node,\footnote{For practical purposes we use only an approximation of the time of passage at the ascending node, which is defined relative to the time of periastron passage $t_\mathrm{p}$ by $t_{ \mathrm{asc} x} = t_p{}_{ X} - P_{ X} \omega_{x}/2\pi $.} inclination of the orbital plane with respect to the plane of the sky, and longitude of ascending node of bodies (or effectively a centre of mass) $x \in \rm\{ p, b, t\}$ and binaries $ X \in \rm \{I,O,\Pi\}$. We note that we use lower-case letters when an orbital element refers to a body in particular, while we use an upper-case letter when the orbital element is associated with the binary as a whole independently of its components. On the first three lines, the correspondence between state vectors and orbital elements is performed using the post-Newtonian formalism of \cite{damour_general_1985} (see also \voisin). Except for the inner binary, this is equivalent to using the usual definition of orbital elements \cite[e.g.][]{beutler_methods_2004}. On the last line, the masses $m_{\rm p,i,o,\pi}$ are related to the mass ratio $m_{\rm i}/m_{\rm p}$ and orbital periods $P_{\rm I,O,\Pi}$ using Kepler's third law. The state vectors of each individual bodies are derived from the left-hand side of Eq. \eqref{eq:orbparam} using centre-of-mass relations at first post-Newtonian order \voisin (the centre of mass of the whole system being at the origin of coordinates, by definition). 

In practice, the fitted parameters are combinations of the orbital elements in Eq. \eqref{eq:orbparam}. These choices reflect usual practices in pulsar timing, as well as relative sensitivity and correlations between parameters (see Tables \ref{tab:fitresults}  and \ref{tab:modelspecific}. Of particular interest are the use of Laplace-Lagrange parameters $e_{ X}\cos\omega_{ X}, e_{ X}\sin\omega_{ X}$ particularly adapted to low-eccentricities \citep[e.g.][]{lange_precision_2001} and the use of $\delta i = i_{\rm I} - i_{\rm O}, \delta \Omega=\Omega_{\rm b} - \Omega_{\rm p}$, which allowed us to better reflect the fact that the two orbits ($\rm I,O$) are co-planar within error bars (see below).

\begin{table*}
	\caption{\label{tab:modelspecific} Mean values of model-specific parameters of the MCMC runs assuming the PL3 (left) and Planet (right) models with their $68\%$ confidence intervals.}
	\centering
	\renewcommand{\arraystretch}{1.15}
	\begin{footnotesize}
		\begin{tabular}{lcl|rcr}
			Parameter & Symbol & PL3GR & PlanetGR & Symbol & Parameter
			% Value \MK{MORE READABLE TO MOVE UNITS TO LEFT COLUMN}: PF- this has been addressed. Table now is in a format more similar to current ways of displaying timing parameters in the literature
			\\
			\hline
			\multicolumn{6}{c}{Fitted values} \\
			\hline
			Fundamental frequency ($\rm day^{-1}$) & $\nu$   &   $3.44(28)_{-54}^{+61} \times 10^{-4}$   &   $3310_{-260}^{+280} $   &   $P_{\rm \Pi}$ & Orbital period ($\mathrm{d}$) \\ 
			Amplitude at $\nu$ ($\rm\mu s$) & $A_\nu$   &   $4.4(82)_{-19}^{+22}$   &   $(6.5)_{-1.4}^{+1.6} \times 10^{-6}$   &   $a_{\rm t}\sin i_{\rm \Pi}$ & Projected semi-major axis (lt-s) \\ 
			Power-law index & $\gamma$   &   $2.73(26)_{-61}^{+68}$   &   $-(1.5)_{-1.6}^{+1.4} \times 10^{-5}$   &   $a_{\rm t}\cos i_{\rm \Pi}$ & Co-projected semi-major axis (lt-s)\\ 
			Phase 1 (rad) & $\phi_1$   &   $-0.21(54)_{-17}^{+19}$   &   $0.2(51)_{-37}^{+32}$   &   $e_{\rm \Pi}\sin \omega_{\rm t}$ & Laplace-Lagrange\\ 
			Phase 2 (rad) & $\phi_2$   &   $-0.77(74)_{-61}^{+68}$   &   $(5.7)_{-2.4}^{+2.4} \times 10^{-2}$   &   $e_{\rm \Pi}\cos\omega_{\rm t}$ & Laplace-Lagrange\\ 
			Phase 3 (rad) & $\phi_3$   &   $-1.42(21)_{-61}^{+68}$   &   $56549_{-20}^{+3333}$   &   ${t_{\mathrm{asc}}}_{\rm t}$ & Time of ascending node ($\mathrm{MJD}$)\\ 
			&    &      &   $1.(25)_{-53}^{+62} \times 10^{2}$   &   $\Omega_{\rm t}$ & Longitude of planet ascending node ($^{\circ}$)\\ 
			\hline
			\multicolumn{6}{c}{Derived values} \\
			\hline
			Fundamental period (days) & $1/\nu$   &   $2904.6_{-5.1}^{+4.6} $   &   $(1.65)_{-0.89}^{+1.6} \times 10^{-5}$   &   $a_{\rm \pi}$ & Semi-major axis (lt-s)\\ 
			Amplitude at $1\rm yr^{-1}$ ($\rm\mu s$) & $A_{1\rm yr^{-1}}$   &   $1.55(16)_{-68}^{+61} \times 10^{-2}$   &   $157_{-54}^{+11} $   &   $i_{\rm \Pi}$ & Orbital inclination w.r.t. plane of sky ($^{\circ}$)\\
			&    &      &   $119_{-42}^{+16} \times 10^{2}$   &   $\delta i_{\rm \Pi}$ & Orbital inclination w.r.t. triple system  ($^{\circ}$) \\ 
			&    &      &   $0.2(57)_{-31}^{+29}$   &   $e_{\rm \Pi}$ & Orbital eccentricity\\ 
			&    &      &   $77.2_{-7.2}^{+6.0}$   &   $\omega_{\rm t}$ & Longitude of periastron  ($^{\circ}$)\\ 
			&    &      &   $57258_{-77}^{+3270} \times 10^{4}$   &   ${t_p}_{\rm \Pi}$ & Time of periastron passage ($\mathrm{MJD}$)\\ 
			&    &      &   $(1.23)_{-0.66}^{+1.1} \times 10^{-8}$   &   $m_{\rm \pi}$  & Planet-companion mass ($\Msol$) 
		\end{tabular}
	\end{footnotesize}
	\tablefoot{ Uncertainties apply to the digits between parenthesis and delimit the 68\% median confidence region, namely the interval excluding 16\% of the distribution above and below it. The central value quoted is the best-fit value. The upper-case index $\rm \Pi$ refers to the planet binary. The lower-case index $\rm \pi$ refers to the planet, and $\rm t$ refers to the triple-system centre of mass. Data spans MJD $56492 - 59480$. Solar System ephemeris is DE430. Parameters common to both models are reported in Table \ref{tab:fitresults}.}
\end{table*}

\subsubsection{Effects of mutual interactions}

Anticipating the results of Sect. \ref{sec:results}, the only measurable effect of a small planet is through the induced R\o mer delay. This results from the fact that its orbit is not relativistic, and its small mass and large orbital inclination suppress the Einstein or Shapiro delay down to the nanosecond level at most. The R\o mer delay is caused by the variation of the projected distance of the pulsar onto the line of sight of an observer located at the Solar System barycentre,
\begin{equation}
\label{eq:roemermutual}
    \Delta_R = -\frac{\vn_\odot\cdot \vR_{\rm p}}{c} =  -\frac{1}{c}\left(\vn_\odot\cdot \vR_{\rm p/b} + \vn_\odot\cdot \vR_{\rm b/t} +\vn_\odot\cdot \vR_{\rm t}\right),
\end{equation}
where $\vn_\odot$ is the unit vector pointing from the Solar System to the pulsar system barycentre, and the right-hand side is obtained using Eq. \eqref{eq:decomporp}. 

In the present analysis, $\vR_{\rm p}$ is computed numerically. However, it is interesting to use the decomposition of Eq. \eqref{eq:decomporp} in order to estimate what can be measured. We show in Appendix \ref{sec:pertubative} that the three terms in Eq. \eqref{eq:decomporp} map to Jacobi coordinates. These coordinates permit a perturbative treatment of the orbital dynamics, which shows that, as said above, each term follows a Keplerian motion at leading order. That means that in first approximation, each term in Eq. \eqref{eq:roemermutual} can be expressed using the usual binary expression of the R\o mer delay (see Eq. \eqref{eq:roepert} or \cite[e.g][]{hobbs_tempo2_2006, lyne_pulsar_2012}). However, the sole detection of the Keplerian R\o mer delay only allows one to constrain five parameters out of seven associated with the planetary binary: the orbital period $P_{ X}$, the semi-major axis projected onto the line of sight $a_{ x} \sin i_{ X}$, the orbital eccentricity $e_{ X}$ and argument of periastron $\omega_{ X}$, where $ x$ and $ X$ are defined as above. 

As per our parametrisation, the mass of the planet is the solution of Kepler's third law:
\begin{equation}
    a_{\rm t}^3 n_{\rm \Pi}^2 = G \frac{m_{\rm \pi}^3}{(m_{\rm t} + m_{\rm \pi})^2},
\end{equation}
where $n_{\rm \Pi} = 2\pi/P_{\rm \Pi}$ is the planet's mean motion. The three stellar masses, and therefore the mass $m_{\rm  t} = m_{\rm  p} + m_{\rm  i} + m_{\rm  o}$, can be determined thanks to mutual interactions, Shapiro and Einstein delays within the triple system \voisinp. However, $a_{ x}$ is degenerate with $\sin i_{ X}$ at Keplerian order. 

In Appendix \ref{sec:pertubative}, we derive a simplified model of the first-order correction to the orbital motion of the planetary binary using Laplace-Lagrange perturbation theory. It amounts to decomposing $\vR_{\rm t} = \vR_{\rm t}^{\rm K} + \delta \vR_{\rm t}$ where the first term describes a Keplerian orbit of the planet with a point mass $m_{\rm t}$ at $\rm t$, and the second term accounts for the finite size of the triple system to first order in the ratio between the size of the planetary binary and that of the outer binary. It leads to a R\o mer delay term $ \delta\Delta_R = -\vn_\odot\cdot \delta\vR_{\rm t}/c$ of the form, Eq. \eqref{eq:delta1},
\begin{equation}
\label{eq:dRmutual}
    \delta\Delta_R = \alpha t \left(\gamma_c\cos n_{\rm O} t + \gamma_s \sin n_{\rm O} t\right),
\end{equation}
where $n_{\rm O} = 2\pi/P_{\rm O}$ is the mean motion of the outer binary and $\alpha, \gamma_c,\gamma_s$ are constants defined in \eqref{eq:alpha}-\eqref{eq:gammac}. This functional time dependence, a sinusoidal oscillation at the outer-binary period the amplitude of which is growing linearly in time, is not degenerate with other components of the timing formula and can in principle allow for the separate measurement of $\alpha\gamma_c$ and $\alpha\gamma_s$. 

Importantly, $\gamma_c$ and $\gamma_s$ are independent functions of inclination $i_{\rm \Pi}$ and longitude of ascending node $\Omega_{\rm t}$, while $\alpha$ is derived from these as well as the Keplerian parameters of the system. This means that it is sufficient to detect the leading order correction of Eq. \eqref{eq:dRmutual} to constrain all seven parameters associated with the planet's orbit and mass.  

\subsection{Test of the equivalence principle}
Each of the above two model categories is also divided in two depending on whether GR is assumed to be correct or if GWEP is being tested.
The equivalence principle can be parametrised at Newtonian and post-Newtonian order by an effective, body-dependent, gravitational constant characterising the interaction between two bodies $a$ and $b$, that is $G_{ab} =G_{ba} = G \left(1 + \Delta_{ab}\right)$, where $G$ is the Newtonian gravitational constant (see also Sect. \ref{sec:intro} and Eq. (20) of \voisin). In GR, $\Delta_{ab} = 0$.

Solar-system experiments put strong constraints on the weak equivalence principle, as well as the strong equivalence principle in the weak-field regime \citep{touboul_space_2019, hofmann_relativistic_2018}. As shown in \voisin, we can use these constraints to assume that within the precision of our experiment and (at least) within the framework of Bergmann-Wagonner scalar-tensor theories, any detectable violation of SEP can only be due to the pulsar, which is the only strongly self-gravitating object. Thus, the only sensitive parameter is $\Delta \equiv \Delta_{\rm pb}$, where $\rm b \in \{i,o,\pi\}$.

%----------------------------------------

\section{Results\label{sec:results}}

We have compared seven different models all assuming that GR is correct. They are summarised in table \ref{tab:modelcomp}. We ran two additional models testing for SEP violation (see below). We inferred the posterior distribution functions (PDF) of each model using a Markov Chain Monte Carlo algorithm (MCMC) following the same procedure as in \voisin. 
In particular, we used the same astrometric priors as in \voisin derived from Gaia DR2 \citep[][]{lindegren_gaia_2018} and spectroscopic observations of the inner white dwarf \citep{ransom_millisecond_2014, kaplan_spectroscopy_2014}. We used our own implementation \voisinp of the affine-invariant ensemble-sampling algorithm described in \cite{goodman_ensemble_2010, foreman-mackey_emcee_2013}. 

The analysis was carried out on a chain sample with a length of at least 60,000, corresponding to more than 100 ensemble autocorrelation times. We used a set of 192 walkers saved to the chain every five iterations of the stretch move \citep[see][]{goodman_ensemble_2010}. We evaluated the accuracy of derived statistics \citep[e.g.][]{dunkley_fast_2005} by estimating the standard deviation of the mean and standard-deviation estimators, $\hat\sigma_{\hat m}$ and $\hat\sigma_{\hat \sigma}$ respectively, over at least 50 sub-chains (each spanning at least two ensemble autocorrelation times). The ratios of these quantities to the full-chain standard deviation $\sigma$, $\hat\sigma_{\hat m}/\sigma$ and $\hat\sigma_{\hat \sigma}/\sigma$ respectively, are used as `convergence ratios'\citep[][]{dunkley_fast_2005}\footnote{We note that \citep[][]{dunkley_fast_2005} estimate directly the variance of the mean of the whole chain using a more refined spectral method while we estimate variances of a sample of sub-chains by the usual variance estimator. These sub-chain variance estimates ought to be larger than full-chain variances since the estimator scales as $\sim 1/n$ where $n$ is sample size. Therefore, albeit simpler, this method is conservative.}. We considered convergence was reached when both indicators were below 0.1 for every parameter chain.

\subsection{Overall model comparison \label{sec:modcomp}}
\begin{table}[h]
    \centering
\caption{Best-fit statistical properties of the models fitted to the dataset of this paper.   \label{tab:modelcomp}}
\begin{tabular}{lllllll}
Model & Npar  &  $\chi^2$  & $R_{\chi^2}$  &     $\Delta$AIC        &            $\Delta$BIC\\
\hline
Planet  &  32  &  +2.4  &  +0.0003  &  +4.4  &  +12  \\
Kepler &  30  &  +4.0  &  0.0002  &  +2.0  &  -5.4   \\
PL2/Kepl1  &  30  &  +19  &  +0.0014  &  +17  &  +9.6  \\
{PL3}  &  {31}  & { 15511.9}  & { 1.2466 } &  {15573.9  }&  {15804.3}  \\
PL4  &  32  &  +4.8  &  +0.0005  &  +6.8  &  +14  \\
PL5  &  33  &  -5.4  &  -0.0002  &  -1.4  &  +14  \\
PL3DM10 &  39  &  -17  &  -0.0006  &  -1.3  &  +58
    \end{tabular}
\tablefoot{Npar is the number of parameters of the model, $\chi^2$ the minimum chi squared, $R_{\chi^2}$ is equal to $\chi^2/\rm Ndof$ where  $\mathrm{Ndof} = 12474 - \mathrm{Npar}$. AIC and BIC are the Akaike and Bayesian information criteria, respectively. The `PL3' line is the reference fit subtracted from all the other lines. All models assume GR. Models are described in Sec. \ref{sec:modcomp}: `Planet' is the full 4-body numerical model; `Kepler' is the purely Keplerian model without mutual interactions; `PLn' are models with $n$ sinusoidal harmonics following Eq. \eqref{eq:RN} in order to account for red noise; PL2 is equivalent to a planet in Keplerian orbit with the centre of mass of the triple system at leading order in eccentricity (no mutual interactions) and thus denoted `PL2/Kepl1'; PL3DM10 fits separate DM within ten evenly distributed intervals on top of three harmonics (We note the other models have DM and DM' parameters, which have been removed here).}
\end{table}

We compared the best-fitting solutions of each model using both the Akaike (AIC) \citep{akaike_new_1974} and Bayesian Information Criteria (BIC) \citep{schwarz_estimating_1978} (see Table \ref{tab:modelcomp}). To do so we computed the best-fitting solutions of each model by fitting them to the data using the deterministic minimizer \texttt{Minuit} \citep{james_minuit-system_1975} starting from the best-fitting solution derived from the MCMC run. The obtained solution was always very close to the MCMC one, as expected, but important compared to the modest differences we found between models. 

We ran four PL$n$ models, from $n=2$ to $n=5$ Fourier components, initialising the PL$n+1$ run from the outcome of the PL$n$ run in order to limit computational cost. We ran two planetary models: a fully Keplerian one (called `Kepler'), and a numerical integration (`Planet')\footnote{We use `Planet' with capital P whenever referring to the model.} including all the effects of mutual interactions. We note that the PL2 model is equivalent to the Kepler model at leading order in eccentricity (see Appendix \ref{apsec:roemerecc2}), which is why we denote it `PL2/Kepl1'. Given the small difference between Kepler and PL2/Kepl1, no additional MCMC was done for Kepler, and its best-fitting solutions are based on a \texttt{Minuit} refit of PL2/Kepl1, thus saving on computing power.

\begin{figure}
	\begin{center}
		\includegraphics[width=0.5\textwidth]{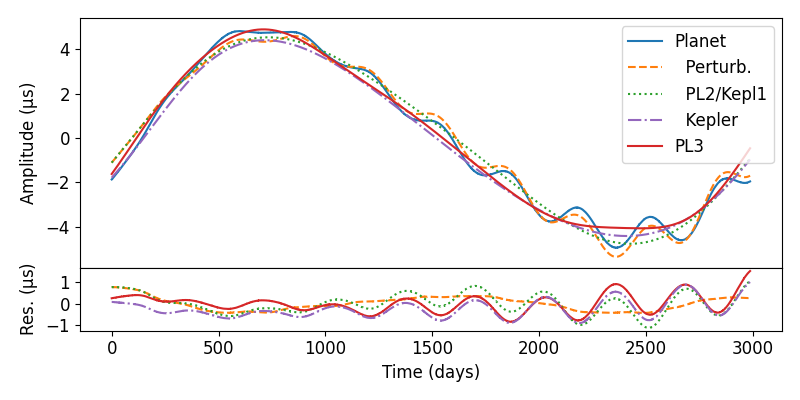}
		\includegraphics[width=0.5\textwidth]{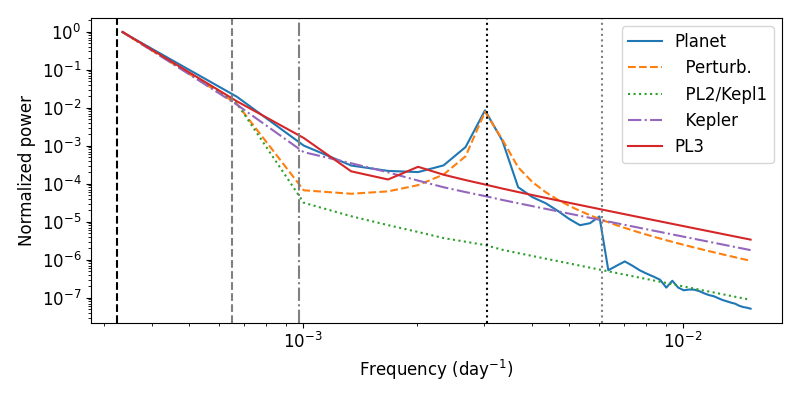}
	\end{center}
	\caption{Comparison of numerically computed R\o mer delay with first-order and Keplerian-order approximations. Approximate expressions have been least-square fitted to the numerical result. Top: Three versions as well as residuals of the least-square fit (lower panel). Bottom: Lomb-Scargle periodogram of the three versions. Vertical dashed lines mark the fundamental (black), second harmonic (grey), and third harmonic (dash-dotted grey) of the planet period. Vertical dotted lines mark the fundamental (black) and first harmonic (grey) of the outer-binary period.\label{fig:comparison}}
\end{figure}

Assuming the presence of a planet, the PL2/Kepl1 and Kepler models differ mostly by the addition of a third harmonic corresponding to the second-order term in eccentricity of the Keplerian R\o mer delay (see Appendix \ref{apsec:roemerecc2}). Contrary to the PL3 model, the phase of the third harmonic is not independent from the two others, and its amplitude is not determined following a power law. Given $e\sim 0.25$ and a projected semi-major axis $x\sim \rm 4\mu s$ (see Table \ref{tab:modelspecific}), the amplitude of this harmonic is only about $\bigcirc(xe^2) \sim  0.25 \rm \mu s$. However small, this additional component appears to be captured by the Kepler model as both AIC and BIC are significantly better than for PL2/Kepl1 in Table \ref{tab:modelcomp}. This can be visualised in Fig. \ref{fig:LS3model} where the small but significant peak ($<0.5\%$ probability compared to white noise) at the third harmonic of the PL2/Kepl1 model residuals disappears with the Kepler model (see also Fig. \ref{fig:comparison}). Thanks to its limited number of parameters the Kepler model has the best BIC of all.

The Planet model has a moderately better $\chi^2$ but due to its two additional parameters accounting for inclination and longitude of ascending node, its AIC and BIC are not as good as those of the Kepler model, and even worse than PL2/Kepl1 concerning BIC. This suggests that mutual interactions may be detected, since the $\chi^2$ is better, but only marginally, which is in line with the large uncertainty on the planet mass (see Table \ref{tab:modelspecific}).

%Comparing the planet model with PL2, we see that the planet model is clearly favoured by AIC ($\mathrm{\Delta AIC \simeq -13}$) and marginally disfavoured by BIC ($\mathrm{\Delta BIC = +2.3}$). This difference between the two criteria is explained by the fact that BIC imposes a stronger penalty for additional parameters. Thus, in the planet hypothesis, it appears that mutual interactions may be marginally detectable and therefore the planet inclination, longitude of ascending node and mass constrained.  

Assuming the red-noise hypothesis, that is the PL$n$ models, PL3 is clearly favoured by BIC and only marginally worse than PL5 according to AIC ($\mathrm{\Delta AIC \simeq -1.38}$). Thus, we make it our reference achromatic red-noise model in the following. PL3 has the best $\chi^2$ of all models, moderately better than Planet ($\Delta \chi^2 \simeq -2.4$), and a moderately better AIC than Kepler ($\Delta \mathrm{AIC} \simeq -2.0$) but significantly worse BIC ($\Delta \mathrm{BIC} \simeq +5.4$).

In Table \ref{tab:modelcomp}, one may notice that PL4 is actually a worse fit than both PL3 and PL5. In theory, since PL3 is nested in PL4, the $\chi^2$ of the latter is expected to be at least as good as that of the former. We interpret the fact that it is not the case here as an occurrence of one of the local minima, which we routinely encountered while fitting. More details are given in Appendix \ref{ap:fitting}, but it is sufficient to say here that although Table \ref{tab:modelcomp} is the result of our best efforts within the limits of reasonable computing resources, it cannot be excluded that the differences between the models are below the level of systematics induced by local minima.

\begin{figure}
\begin{center}
    \includegraphics[width=0.5\textwidth]{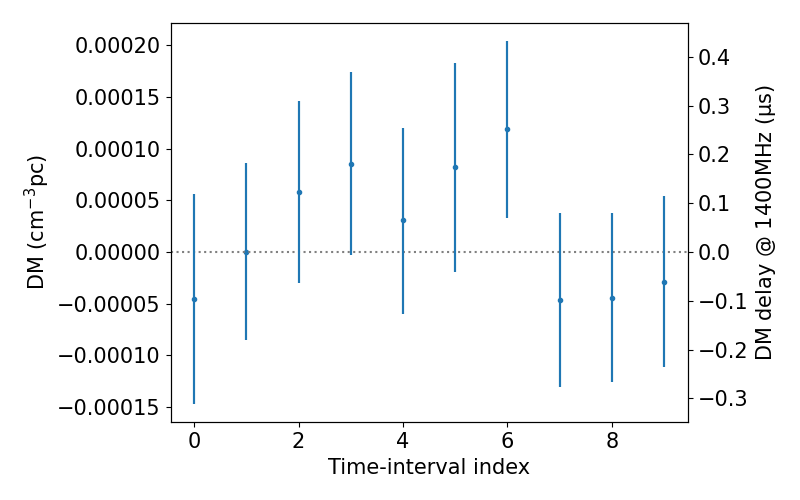}
\end{center}
\caption{Dispersion measure per time interval in the model PL3DM10. The x axis gives the time interval index, and the intervals are equal. Error bars delimit the 68\% confidence region. \label{fig:DMPL3}}
\end{figure}

We also ran a model called PL3DM10, for which the MCMC only sampled the six achromatic red-noise parameters, the ten $\rm DM_X$ DM parameters as well as the pulsar spin parameters $\bar f, \bar \dot f$ (see table \ref{tab:fitresults}) since they may also correlate with the long-term effects of red noise. All orbital and astrometric parameters were fixed to their best-fitting PL3 values so as to optimise computational costs. We note that DM and $\rm DM'$ were also fixed to their PL3 value, thus subtracting that linear trend from the computed $\rm DM_X$. The resulting DM values are shown in Fig. \ref{fig:DMPL3} as a function of the corresponding time interval. No clear DM variation is seen given that the dispersion of the $\rm DM_X$ values is comparable to their uncertainties, and to the uncertainty on the global DM parameter as well (Table \ref{tab:fitresults}). However, we note that AIC marginally favours PL3DM10 over PL3 ($\rm \Delta AIC = -1.3$) but that BIC strongly rejects it ($\rm \Delta BIC = 58$) owing to the large number of additional parameters required. 

% Comparison between planet and PL3
\subsection{Details of the Planet and PL3 models}
\label{sec:resdetails}

%%% 
% Position of Fig.2 before moving to additional material "residuals-PL3-difPL3_planet"
%%%%%

The inherently different nature of the Planet and PL3 models justify a more detailed look at both of them. If the planet hypothesis is restricted to the Kepler model, which can be seen as a sub-model of Planet, then it performs similarly to PL3 across all statistical metrics in Table \ref{tab:modelcomp}. In addition, the potential systematic errors in best-fit estimations do not reasonably allow us to favour one hypothesis over another solely based on the statistical criterion (Appendix \ref{ap:fitting}).

We have collated the results of the MCMC Planet and PL3 runs in Table \ref{tab:fitresults} for parameters common to the two models and Table \ref{tab:modelspecific} for model specific parameters. Additionally partial correlation plots of each PDF are given in Appendix \ref{ap:correlplots}, and complete ones can be found on the repository in footnote \ref{fn:zenodo}. 

The timing signal introduced by both models are shown in Fig. \ref{fig:comparison}. In that figure we have also represented the signal from the PL2/Kepl1 model, the Kepler model (Appendix \ref{apsec:roemerecc2}), as well as the perturbative model described in Appendix \ref{sec:pertubative}. It can be seen that the Kepler model captures the third orbital-frequency harmonic similarly to PL3, while the perturbative model captures the dominant contribution from mutual interactions but lacks power at the third harmonic due to its limitation to first order in eccentricity. In the frequency domain, one sees that the largest difference between the PL3 and Planet models peaks at the outer binary frequency, as expected from the perturbative model, and decays in lower and upper frequency tails around that peak. However, as can be seen in the periodogram of best-fit residuals in Fig. \ref{fig:LS3model}, this particular frequency does not allow for a clear signature with the current data as it is already very well fitted due to the outer binary contribution. In the case of PL3, Kepler, and PL2/Kepl1, it even appears to be over-fitted. Besides that, both the Planet, Kepler, and PL3 models appear consistent with white-noise in Fig. \ref{fig:LS3model} insofar as their residuals lie within the 95\% confidence region for such noise.

In Table \ref{tab:fitresults}, parameters that are common to both models are equal within error bars. Uncertainties on orbital parameters appear to be similar although systematically slightly smaller in the Planet model. On the other hand, uncertainties of spin parameters are an order of magnitude smaller in the PL3 model. These two parameters are the only two that significantly correlate with the low-frequency signal due to their ability to effectively absorb parabolic trends. Therefore, this difference probably comes from the two different ways that spin parameters correlate with model-specific parameters. 

It is noteworthy that the orbital plane of the inner binary still cannot be separated from that of the outer binary, which is reflected by their relative inclination $\delta i$ and relative longitude of ascending nodes $\delta \Omega$ being consistent with zero within error bars in both models. 
We also note that the tension in declination proper motion $\mu_\delta$ compared to its GAIA prior that was observed in \voisin has disappeared in the present results.

The longitude of ascending node $\Omega_{\rm b}$ is incompatible with that of \voisin. During the course of the MCMC runs (irrespective of the models), a peak at the beat frequency between the Earth orbital period and outer-binary period became clearly visible in the residual periodogram as the low-frequency signal was being fitted out. This peak is characteristic of the annual-orbital parallax \citep{kopeikin_possible_1995}, which in the absence of mutual interaction is the only way to measure $\Omega_\mathrm{b}$ by timing.\footnote{It is also possible to use scintillation arcs \citep[][]{reardon_precision_2020}.} However, the MCMC seemed unable to find a solution accounting for this peak while staying in the vicinity of the initial value of $\Omega_\mathrm{b}$, and the assumption was made of a local minimum around the $\Omega_{\rm b}$ value from \voisin. The MCMC chain was restarted with values of $\Omega_b$ offset by $\pm 90\deg$ and $180\deg$ from that of \voisin. All converged to the value reported in Table \ref{tab:fitresults}, which successfully fits out annual-orbital parallax. This value is $180\deg$ away from \voisin (within uncertainty), thus suggesting that mutual interactions in the triple system can constrain the longitude of ascending node with a degeneracy of $\pm 180\deg$ lifted by the detection of annual-orbital parallax. Dedicated analytical or numerical work is needed to verify this conjecture. 

An important point is that the uncertainty re-scaling parameter EFAC is now down to $\simeq 1.11$ from $\simeq 1.31$ in \voisin, which translates into a significant improvement in accuracy. This is due to the fact that EFAC was previously absorbing the dispersion resulting from the unaccounted low-frequency signal. On the other hand, a fit with \texttt{Minuit}  \citep{james_minuit-system_1975} indicates that EFAC would need to be up to $\sim 1.6$ in order to accommodate that signal in the longer dataset of this paper if it remained unmodelled

\subsection{Galactic motion}\label{subsec:gal_motion}

\begin{figure}[ht]
\begin{center}
\includegraphics[width=0.45\textwidth]{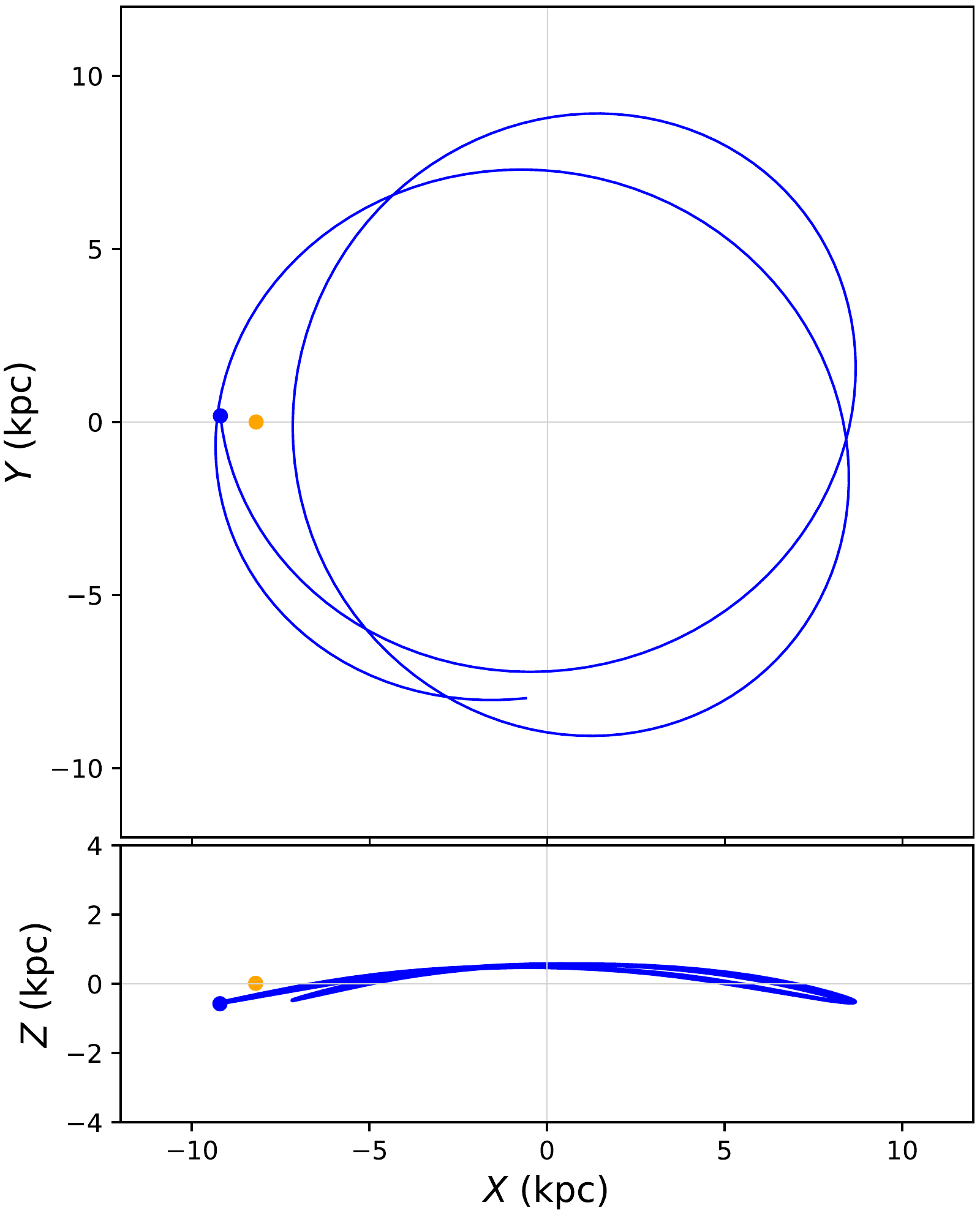}
\end{center}
\caption{Galactic motion of the J0337 system during the past 500~Myr. The blue dot marks its current position. The orange dot shows the location of the Sun. The orbit was calculated with the Galactic gravitational potential and the software provided by \cite{McMillan_2017}. \label{fig:GalMotion}}
\end{figure}

\begin{figure}[ht]
\begin{center}
\includegraphics[width=0.45\textwidth]{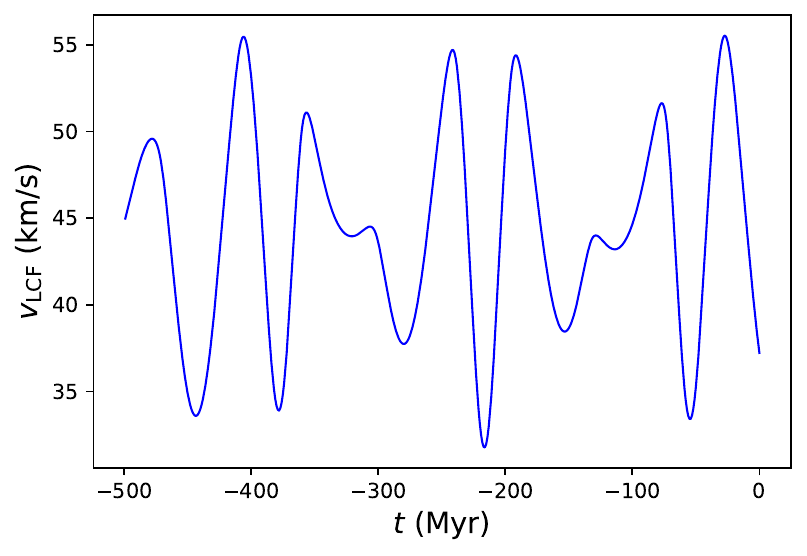}
\end{center}
\caption{Velocity of the J0337 system with respect to the local galactic co-rotating frame during the last 500~Myr. \label{fig:V_LCF}}
\end{figure}

For J0337 we are in the rare position of knowing the 3D velocity of a pulsar system with respect to the Solar System barycentre. The transverse velocity (magnitude and direction) is obtained from proper motion and distance (see Table~\ref{tab:fitresults}). High resolution spectroscopy of the inner white dwarf allowed for the determination of the systemic radial velocity: $29.7 \pm 0.3~{\rm km\,s^{-1}}$ \citep{kaplan_spectroscopy_2014}. With this information at hand, starting with the current location of the pulsar one can integrate its Galactic motion back in time. To do this, we use the Solar position and velocity parameters, the Galactic potential, and the software of \cite{McMillan_2017}. Figure~\ref{fig:GalMotion} shows the Galactic orbit of the J0337 system during the past 500~Myr. It is obvious that the system follows approximately the overall Galactic rotation. Indeed, its speed relative to the frame co-rotating with the Galaxy remains in the range $\sim [30, 55]~\rm km\, s^{-1}$ in the entire integration span (see Fig.~\ref{fig:V_LCF}). This finding is of particular importance for Sec.~\ref{sec:formation}, where we discuss the evolutionary history of the system, as it suggests that the formation of the pulsar had only a small kick imparted on the system.

% SEP runs 
\subsection{Tests of the strong equivalence principle with the Planet or PL3 models}
\label{sec:SEP}

\begin{figure}
\begin{center}
    \includegraphics[width=0.5\textwidth]{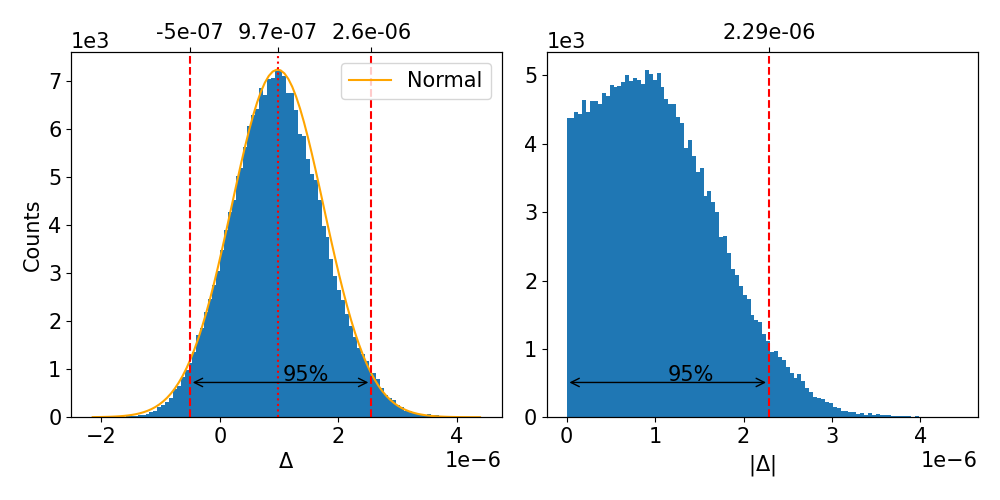}
    \includegraphics[width=0.5\textwidth]{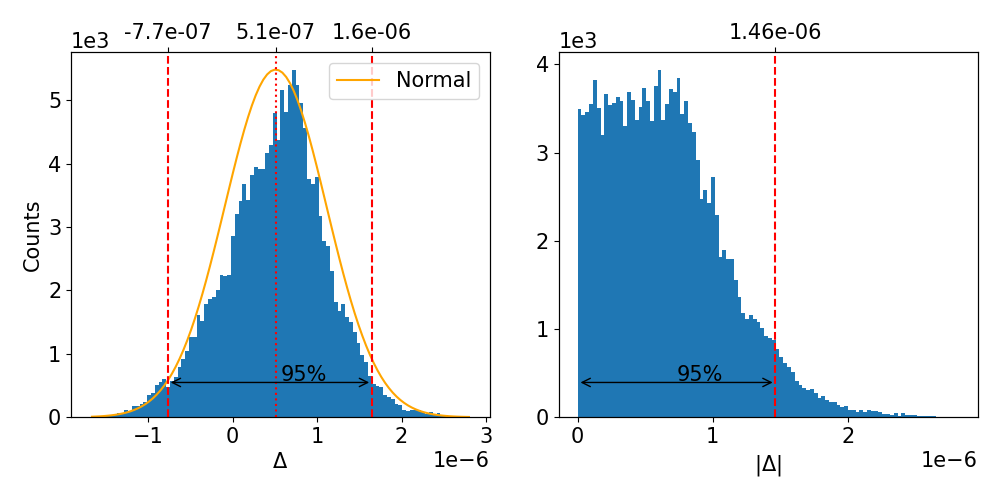}
\end{center}
\caption{Measurements of the SEP violation parameter $\Delta$ (left column) and its absolute value $|\Delta|$ (right column) assuming the PL3 (upper row) or Planet (bottom row) models. Vertical dotted lines mark the mean value of the distributions, while the vertical dashed lines delimit the 95\% confidence regions.
\label{fig:SEP}}
\end{figure}

In order to test the SEP, we completed MCMC runs of the PL3 and Planet models while letting the SEP $\Delta$ parameter free (instead of fixed $\Delta=0$ when assuming GR). For other parameters results are compatible with the GR runs reported in Table \ref{tab:fitresults} and \ref{tab:modelspecific}, although with much wider error bars for the orbital parameters due to their strong correlations with $\Delta$. Besides the extended dataset and lower EFAC, this mainly explains why uncertainties reported in Table \ref{tab:fitresults} can be orders of magnitudes smaller than in \voisin, which reported only an SEP run. In Fig. \ref{fig:SEP}, we report the marginal PDF of the SEP $\Delta$ parameter for both models. Their 95\% confidence region can be expressed as 
\begin{eqnarray}
\Delta = 0.97^{+1.58}_{-1.48}\times 10^{-6} &\text{ and }& |\Delta|  <  2.29\times 10^{-6} \; \text{ (PL3)},\\
\Delta = 0.51^{+1.14}_{-1.28}\times 10^{-6} &\text{ and }& |\Delta|  <  1.46\times 10^{-6} \; \text{ (Planet)}.
\end{eqnarray}
In both cases the width of the confidence region of $\Delta$ is reduced by $\sim 12\%$ with respect to \voisin. This can be related to the reduction by $\sim 16\%$ of the EFAC parameter assuming a locally linear dependence on this parameter. 
However, the bound on $|\Delta|$ is improved by 30\% with respect to \voisin in the Planet model while it worsens by 10\% in the PL3 model. The latter case is due to the fact that the mean value of the distribution is offset from zero by more than one sigma.

\section{Discussion\label{sec:discussion}}

 Our results show that DM variations cannot be responsible for the $\sim 4 \rm \mu s$ amplitude of the observed low-frequency signal (Fig. \ref{fig:DMPL3}), but may at most contribute marginally at the $\sim 0.1 \rm \mu s$ level, if at all. Thus, in what follows we consider only the achromatic models. 
 
\begin{figure}
	\begin{center}
		\includegraphics[width=0.5\textwidth]{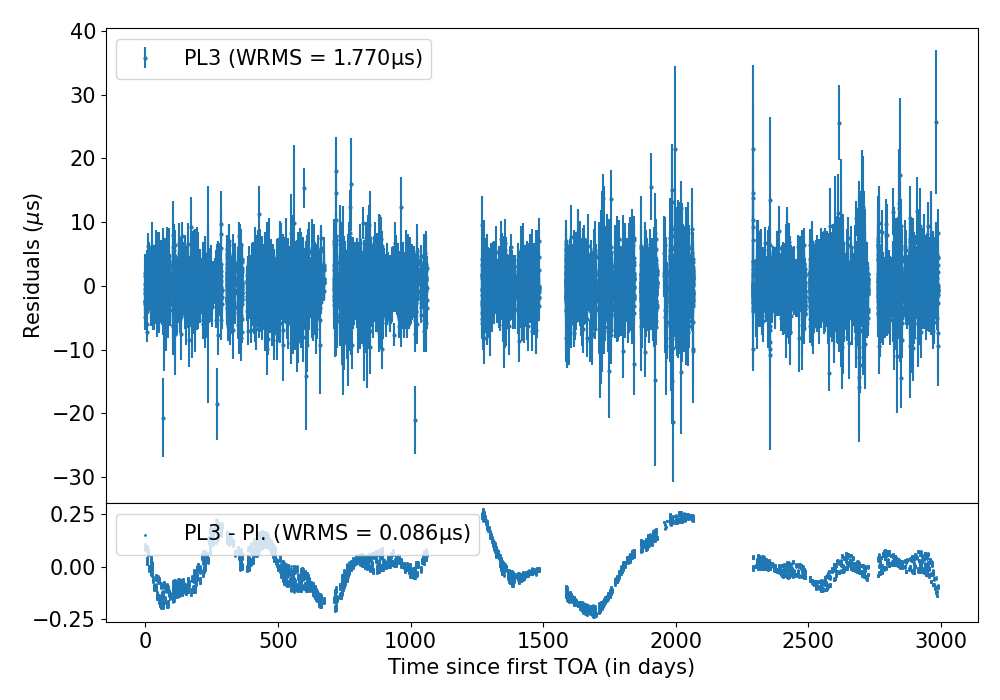}
	\end{center}
	\caption{Timing residuals. Top: Best-fit residuals of model PL3. Bottom: Difference between residuals of best-fit PL3 and Planet models. Error bars are not shown for clarity, but the median 1-sigma uncertainty is $1.9\rm \mu s$, and the mean is $2.2\mu s$.  \label{fig:residualsPL3}}
\end{figure}

The red-noise model following a power-law spectrum with three Fourier components (the PL3 model) provides the best description of the data from an information-theoretic point of view.
However, the physical nature of the signal in the planet and the red-noise hypotheses are completely different, while the difference in terms of fitting residuals is minute, as can be seen in Figs. \ref{fig:LS3model} and \ref{fig:residualsPL3}. Besides, even in the Planet model a moderate red-noise component is not unexpected, typically at the $\sim 0.1 \rm \mu s$ level. We did not try such a planet+red noise model since the additional complexity seems disproportionate with respect to the Planet model residuals, and would likely lead to over-fitting. However, our main goal was to address the large $4 \rm \mu s$ signal, which both red-noise and Planet models do successfully. Thus, in what follows we discuss separately the two hypothesis.

% % Comparison Planet vs PL5
 \subsection{Achromatic red noise hypothesis}
 \label{sec:rndiscussion}

%Red noise
Red noise in millisecond pulsars has been extensively studied in the frame of pulsar timing arrays (PTAs). Here we compare the red-noise properties of J0337 to those of the 67 different pulsars\footnote{The six EPTA pulsars in the cited reference are also among the NANOGrav 47 pulsars, and we count 6 out of 26 pulsars in common with the PPTA dataset.} studied by the NANOGrav \citep{alam_nanograv_2021}, EPTA \citep{chalumeau_noise_2022}, and PPTA \citep{goncharov_identifying_2021, reardon_parkes_2021}. In those works, red noise is modelled as a Gaussian process over a Fourier basis characterised by a power spectrum density of the form $P(\nu) = \alpha A_{\rm GP}^2 (\nu/\rm yr^{-1})^{-\gamma_{\rm GP}}$ where $\alpha = 1/12\pi^2$ in \cite{chalumeau_noise_2022, goncharov_identifying_2021} or $\alpha=1$ in \cite{alam_nanograv_2021}, $A_{\rm GP}$ is the amplitude of the process at frequency $\nu = 1\rm yr^{-1}$ and $\gamma_{\rm GP}$ is its exponent. Although we did not use a Gaussian process framework and therefore cannot make a rigorous comparison, we may estimate the parameters of a Gaussian red-noise process that would produce the same average spectrum as the one we fit in this paper. Up to an order one factor, we get $A_{\rm GP} \sim A \left(T/\alpha\right)^{1/2} \left(\nu/\mathrm{yr^{-1}}\right)^{\gamma_{\rm GP}/2}$ where we used the notations of Eq. \eqref{eq:RN}, $T$ is the time span of observations, and $\gamma_{\rm GP} \sim 2\gamma$. Using the results of our PL3 model, we obtain\footnote{Equivalently $\log_{10}(A_{\rm GP}/\mathrm{yr^{3/2}}) \sim -13.9$ with $\alpha=(12\pi^2)^{-1}$ or $\log_{10}(A_{\rm GP}/\mathrm{yr^{3/2}}) \sim -14.9$ with $\alpha=1$.} $A_{\rm GP} \sim \alpha^{-\frac{1}{2}} 0.04\,\rm \mu s\, yr^{1/2}$ and $\gamma_{\rm GP} \sim 5.4$. 

The value of the exponent places it among the steepest red-noise spectra, as is expected from a very smooth, quasi-sinusoidal signal. On the other hand, the amplitude of the signal, $\sim 4.4 \rm \mu s$ at a period of 8 years ($\sim 4.4 \rm \mu s @ 8\rm yr$), is only surpassed by PSR J1824-2452A with $\sim 8 \rm \mu s @ 8\rm yr$ and $\gamma_\mathrm{GP}\simeq 5$ among pulsars with a similarly steep spectrum \citep{goncharov_identifying_2021}. It is followed by PSR J1939+2134 (PSR B1937+21) with $\sim 1 \rm \mu s @ 8\rm yr$ and $\gamma_\mathrm{GP}\simeq 5.4$ \citep{goncharov_identifying_2021, reardon_parkes_2021}. 
Interestingly, PSR J1824-2452A and PSR J1939+2134 are both somehow extreme MSPs. They have the second and third largest spin-down power of all MSPs, $\dot E \sim 2 \times 10^{36}\rm erg/s$ and $\sim 1\times 10^{36}\rm erg/s$ respectively \citep{reardon_parkes_2021}, as well as the second and fourth largest spin-down rates, according to the ATNF catalogue \citep{manchester_australia_2005}, with $\dot f \sim -2\times 10^{-13} \rm Hz/s$ and $\dot f \sim -4\times 10^{-14} \rm Hz/s$ respectively \cite{reardon_parkes_2021}. They are also among MSPs with the youngest characteristic ages with 30 and 240 million years, respectively. All these quantities are one to two orders of magnitude above the bulk of MSPs. Both pulsars are also among the few MSPs known to produce giant pulses \citep[e.g.][]{knight_study_2006}. These characteristics are significant whether one considers that achromatic red noise is due to activity in the magnetosphere \citep{lyne_switched_2010} or to turbulence in the superfluid core of the neutron star \citep{melatos_pulsar_2014} since in both cases a correlation with spin-down rate is expected. 

Empirically, a correlation is indeed observed over the general pulsar population \citep[][]{hobbs_analysis_2010, lyne_switched_2010, shannon_assessing_2010}. A scaling law $\propto f^\alpha|\dot f|^\beta$ can be fitted to the amplitude of red noise. In particular, \citep[][]{shannon_assessing_2010} found $\alpha\simeq -1.4 ; \beta\simeq 1.1$. At that time only a few MSPs showed detectable noise, notably PSR J1939+2134, and it appeared consistent with that scaling law. Restricted to a population of pulsars with comparable spin frequencies $f$ this law means that noise should correlate with spin-down rate $\dot f$ or even spin-down power. More generally, it indicates that the characteristic spin-down age $\tau = f|\dot f|^{-1}$ can be used as a reasonable estimate, as was also noted in \citep[][]{hobbs_analysis_2010}: the younger the more noisy.

On the other hand, J0337 does not show any uncommon intrinsic properties but a spin-down power of $\dot E\sim 3\times 10^{34}\rm erg/s$, spin-down rate of $\dot f \sim -2\times 10^{-15} \rm Hz/s$, and characteristic age of 2.4 billion years. These quantities place J0337 in the bulk of the distribution of MSPs and according to the aforementioned scaling law its noise amplitude should be at least five to ten times weaker. We note that this is actually the case of the other MSPs with comparably steep index (>4) in \citep[][]{chalumeau_noise_2022, goncharov_identifying_2021, alam_nanograv_2021}.  However limited, these comparisons suggest that if indeed red noise causes the observed signal from J0337 then it is unusually strong, especially when one considers how steep its spectrum is. 

\begin{figure}
\begin{center}
    \includegraphics[width=0.45\textwidth]{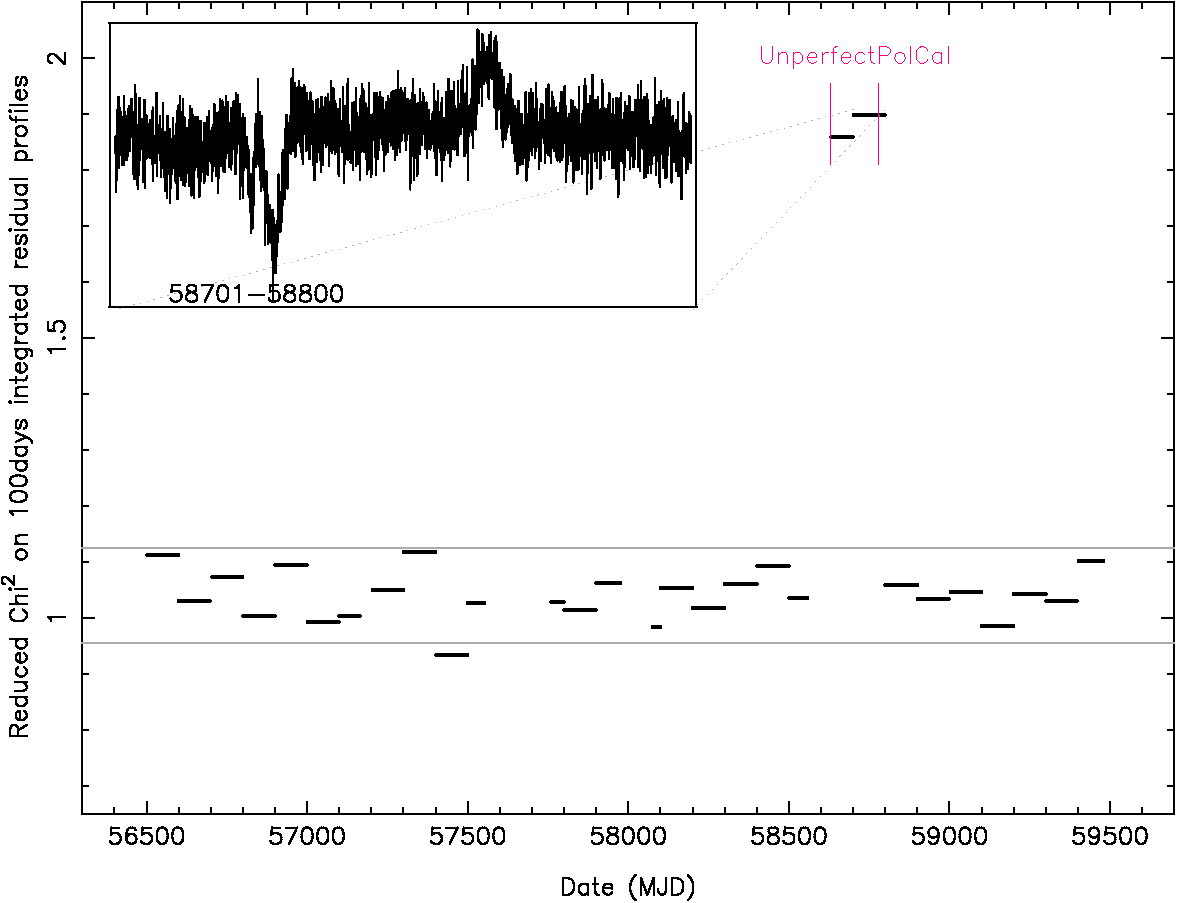}
\end{center}
\caption{Reduced $\chi^{2}$ of the residuals between each 100-day sub-profile and the main template as a function of date. Two sub-profiles are in the time interval when imperfect polarisation calibration was performed (`UnperfectPolCal', see main text). The inset shows the residuals obtained for one of these sub-profiles, while others do not show discernible structure (i.e. noise). The two horizontal lines delimit the two-standard-deviation interval around the mean of the reduced $\chi^{2}$ distribution corresponding to $\sigma_{\rm{red.} \chi^2} \simeq 0.043$ (imperfect polarisation region excluded).   \label{fig:prfvariation}}
\end{figure}

In some of the occurrences where the timing noise is strong, a correlated variation of the pulse profile has been observed \citep{lyne_switched_2010}. Consequently, we have tested the presence of such a slow variation of the pulse shape. To do so, we have built a series of 100-day integrated profiles distributed over the whole observation span. We then used the \texttt{pat} tool of \texttt{PSRCHIVE}  in order to compare them to our main template used for ToA determination after correcting for phase shifts. 
The differences between the main template and sub-integrations can be quantified by calculating their reduced $\chi^{2}$, as shown in Fig.\ref{fig:prfvariation}. The reduced $\chi^{2}$ is usually close to 1, indicating that we do not detect any significant pulse shape variations, except in May-Oct 2019 for $\sim$150 days (MJD 58631 to MJD 58780). This corresponds to the period of instrumental modifications (see Sec. \ref{sec:observations}) that we have eventually excised. In order to quantify the dispersion of reduced $\chi^2$, we show their two-standard-deviation interval in Fig. \ref{fig:prfvariation} computed on the clean $\chi^2$ sample only. All but one element fall into that region, in line with expectations. Indeed, each profile possesses $\sim 2000$ degrees of freedom, thus the $\chi^2$ distribution is well approximated by a Gaussian law and the above-mentioned interval corresponds to a probability of  95\%.

\subsection{Planet hypothesis}
\label{sec:planetdiscussion}
Assuming a mere Keplerian orbit of the planet without mutual interactions, the Kepler model appears to be favoured by BIC and moderately disfavoured by $\chi^2$ and $\rm AIC$. The full Planet model does not improve sufficiently its $\chi^2$ in order to improve either AIC or BIC compared to Kepler, but its two additional parameters are nonetheless constrained by  MCMC albeit with relatively large uncertainties. 
%The full orbital motion with mutual interactions is either favoured over a Keplerian orbit, according to AIC, or at least marginally equivalent, according BIC. 
This allows us to discuss the complete solution constraining all seven parameters reported in Table \ref{tab:modelspecific}, and in particular what they indicate about the formation and stability of the system.

The Planet model points to a very low-mass planet in a wide eccentric orbit around the stellar triple system. The mass of the planet is $1.23^{+1.1}_{-0.66} \times 10^{-8} M_\odot$, which is about $1.3\times 10^{-5} M_{\rm Jup}$, $\sim 0.004 M_{\rm Earth}$ or $\sim 0.4 M_{\rm Moon}$. As such, it is lighter than any of the planets orbiting PSR B1257+12 \citep{konacki_masses_2003} and could be the lightest exoplanet to date according to the extrasolar planet encyclopedia\footnote{The catalogue at \url{exoplanet.eu} reports one lighter planet around the white dwarf WD 1145+017. However, it seems that this object could be composed of multiple disintegrating planetesimals \cite{croll_multiwavelength_2017}. }. 

\begin{figure}
\begin{center}
    \includegraphics[width=0.5\textwidth]{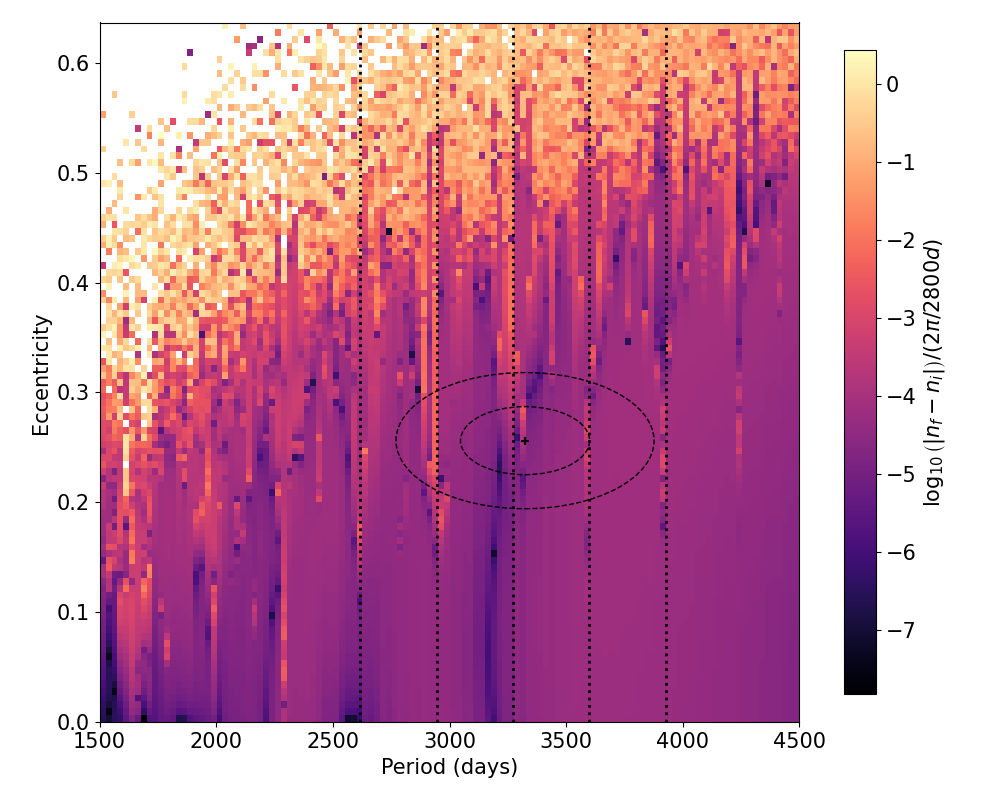}
\end{center}
\caption{Chaos map representing the variation of the proper mean motion $|n_f -n_i|$ between two halves of a 2000-year integration on a grid of $120\times 120$ initial conditions in the period-eccentricity plane of the planet. White pixels represent orbits that are so chaotic that we could not run the frequency analysis. The eccentricity grid initially went all the way to 0.9, but we show here only the 102 lower values, as nearly all pixels in the last 18 lines are white. Vertical dotted lines show the positions of resonances 8:1 to 12:1 with the outer-binary period. The cross shows the mean parameters of the planet as reported in Table \ref{tab:fitresults}, and the ellipses represent the one and two-sigma confidence regions, respectively. \label{fig:chaosmap}}
\end{figure}

We have evaluated the short-term stability of the orbit of the candidate planet by applying the frequency analysis method of \citet{laskar_secular_1988, laskar_chaotic_1990, laskar_frequency_1993, Laskar_2005} in the eccentricity-period plane, Fig. \ref{fig:chaosmap}, with the help of the TRIP software \citep[][]{gastineau_trip_2011}. These two parameters are relevant for stability, as they control the distance to the triple system, and other parameters are fixed to their best-fitting values. 
If the system was perfectly integrable, it would possess an intrinsic set of fundamental frequencies (related to the existence of action-angle coordinates), which would be constant over time. These frequencies could be retrieved very accurately by running the frequency analysis method on a numerical solution of the system. For a weakly chaotic system, however, fundamental frequencies can still be defined, but they are now valid only in a restricted interval of time, as chaos makes the system slowly diffuse in the available region of the phase space. The measure of the drift of the fundamental frequencies then gives a direct indication of the level of chaos present in the system \citep[see][]{laskar_chaotic_1990}. Here, we are interested in the fundamental frequency $n$ of the planet associated with its orbital motion (i.e. its `mean', or `proper' mean motion): its drift rate informs us about the short-term orbital stability of the planet. In practice, we run the frequency analysis on the two halves of a 2000-year numerical integration of the orbits; Fig. \ref{fig:chaosmap} shows the difference between the two values of $n$ obtained.

We can see in Fig. \ref{fig:chaosmap} that larger eccentricities or smaller periods lead to unstable motion due to the stronger influence of the inner triple system. Mean-motion resonances with the outer binary are also a noticeable source of chaos. Due to its large uncertainty, the confidence region of the planet overlaps with the unstable stripes associated with the 9:1 to 11:1 resonances. However, most of the region remains compatible with short-term stability, and it contains niches in which virtually no chaos is detected (values $< 10^{-5})$. Due to the very short period of the inner binary, studying the long-term (Gyr) stability of the system would require building a dedicated averaged model, which is out of the scope of the present work. We have nonetheless pushed numerical integration of the best-fitting solution to 100,000 years with reasonable accuracy (and to 100 Myrs without GR), and we did not see any long-term drift of the planet's orbital elements. 

\begin{figure}
    \includegraphics[width=0.5\textwidth]{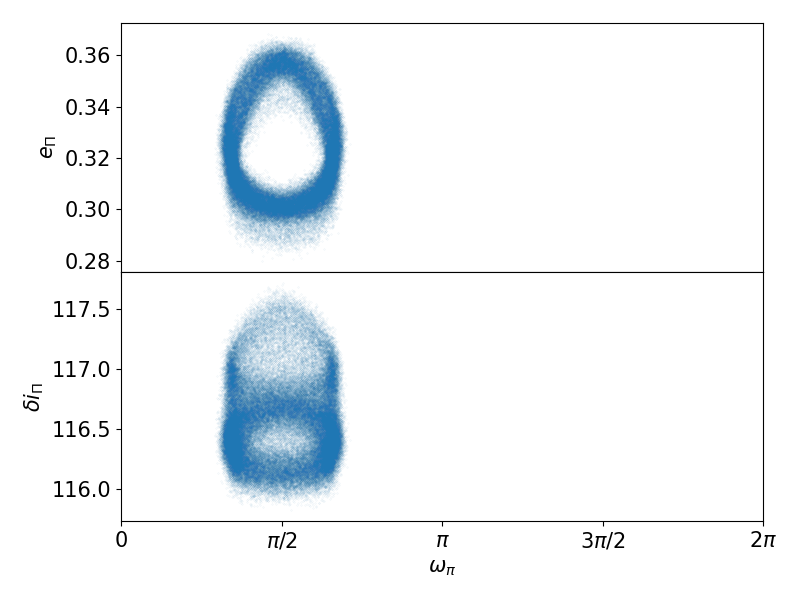}%Kozai-planet_ecc-inc.pdf}
\caption{\label{fig:kozaiei} Evolution of the eccentricity and inclination of the planet as a function of its argument of periastron. Orbital elements are measured with respect to the orbital plane of the inner two binaries. The small dots show the trace of a 20-Myr numerical integration of the planet, as given by an N-body integrator without GR. Initial conditions are chosen near the best-fit orbit.}
\end{figure}

However, long-term numerical integrations of the best-fitting solution reveal an intriguing behaviour for the planet's orbit. First, we note that the best-fit inclination of the planet's orbital plane with respect to the plane of the inner two binaries is $\delta i_{\rm\Pi} = 119^{+16}_{-42}{}^\circ$ (Table \ref{tab:fitresults}). This indicates that the motion of the planet is very inclined, and very likely retrograde, with respect to the orbits of the two inner binaries. More puzzling, this confidence interval is centred on the inclination of the exterior von Zeipel-Lidov-Kozai resonance,\footnote{This is the usually called Lidov-Kozai effect. We follow here \citet{ito_lidov-kozai_2024} who showed that earlier work by von Zeipel should also be acknowledged. See references in the text.} whose location can be approximated by $\cos(\delta i_{\rm\Pi}) = -1/\sqrt{5}$, that is, $\delta i_{\rm\Pi} \simeq 116.6^\circ$ \citep[see][]{gallardo_survey_2012, saillenfest_long-term_2016}. Contrary to the classic von Zeipel-Lidov-Kozai resonance for an outer perturber \citep[see][]{zeipel_sur_1909, lidov_evolution_1962, kozai_secular_1962}, this resonance appears beyond quadrupole order, so its width is quite narrow in inclination. And yet, 
long-term numerical integrations with initial conditions chosen near the best-fit solution show that the planet can stably librate within the resonance. This peculiarity is directly apparent when using a system of coordinates in which the third axis is aligned with the total angular momentum. In this reference frame, the orbital inclination of the two inner binaries is close to zero and the inclination of the planet is close to $\delta i_{\rm \Pi}$. As the planet is in the Kozai resonance, its eccentricity and inclination are affected by correlated oscillations, and its argument of periastron oscillates around $\pi/2$ (instead of circulation between $0$ and $2\pi$ see Fig. \ref{fig:kozaiei}).
As the resonance only spans a few degrees in inclination, obtaining this configuration in a hierarchical system with nearly circular inner perturbers may seem hard to attribute to chance -- even though the confidence region is wider than the resonance.

Considering this unlikely coincidence, we may conjecture that the Kozai resonance stabilises the planet’s orbit, possibly because of the bounded secular evolution of the argument of periastron $\omega_\pi$ it implies. Indeed, an argument of periastron confined near $\pi/2$ means that when the planet is at pericentre, it is always away from the orbital plane of the two binaries. If this conjecture is correct, then the planet may be the only remnant of a larger population of small bodies that were on less stable orbits, or alternatively may have formed there from the gas expelled during the recycling of the pulsar as discussed below.

\subsection{Planet formation and survival}
\label{sec:formation}

The formation of the triple compact-object system J0337 is truly remarkable, and modelling its formation is a major challenge \citep{tauris_formation_2014}. The system has survived at least three stages of mass transfer (including one common envelope (CE) phase) and the supernova (SN) explosion that created the pulsar. On top of this, the system has managed to remain dynamically stable on a long-term (Gyr) timescale. 
\citet{tauris_formation_2014} found a possible solution for J0337 using a self-consistent and semi-analytical approach. Their solution is not unique due to degeneracy, and it should thus be considered {\em a} solution rather than {\em the} solution.
An interesting question is if their formation model can be extended to accommodate the formation, and survival, of a planet?

In the following, we adapt the formation model of \citet{tauris_formation_2014} --- see their Table~1 and Fig.~1 for an overview of the various stages involved (with the corresponding stage numbers, which we refer to below). We begin by noting that in stage~2, it has been suggested that planets may form from the condensation of vast amounts of material ejected in the equatorial region during a CE phase \citep{bhd+10,sd14,bs14}.

Although we cannot rule out the possibility of an outer planet forming from material ejected by the giant-star progenitors of the WDs (stages~6 and 8), such a formation process seems unlikely. It would require a rare and complex mechanism, facing many challenges not present in traditional planet formation models \citep{dbl+23}. One major issue is that, even for a single red-giant star, enough dust and gas would need to accumulate in the circumstellar envelope for planetesimals (the small building blocks of planets) to form. Additionally, the gas and dust would need to cool and condense over a prolonged period before a planet could develop, and this cooling phase is critical since high temperatures would inhibit planet formation. In the formation model of \citet{tauris_formation_2014}, the material would likely be ejected via isotropic re-emission during RLO with high pressure and temperature, meaning the material would likely be blown away before planet formation could occur. That said, it remains a puzzle how some WDs and pulsars have orbiting planets, and further research is required on this topic.

Assuming a planet did form during the CE ejection in stage~2, we can still adopt the original scenario and model parameters derived by \citet{tauris_formation_2014}, given the very small planet mass of only $M_{\rm planet}=1.23\times 10^{-8}\;M_\odot$. At stage~4, just prior to the SN explosion, we can calculate the probability that this planet survived the kinematic effects of the SN.

\begin{figure}
\vspace{-0.8cm}
\begin{center}
    \includegraphics[width=0.55\textwidth]{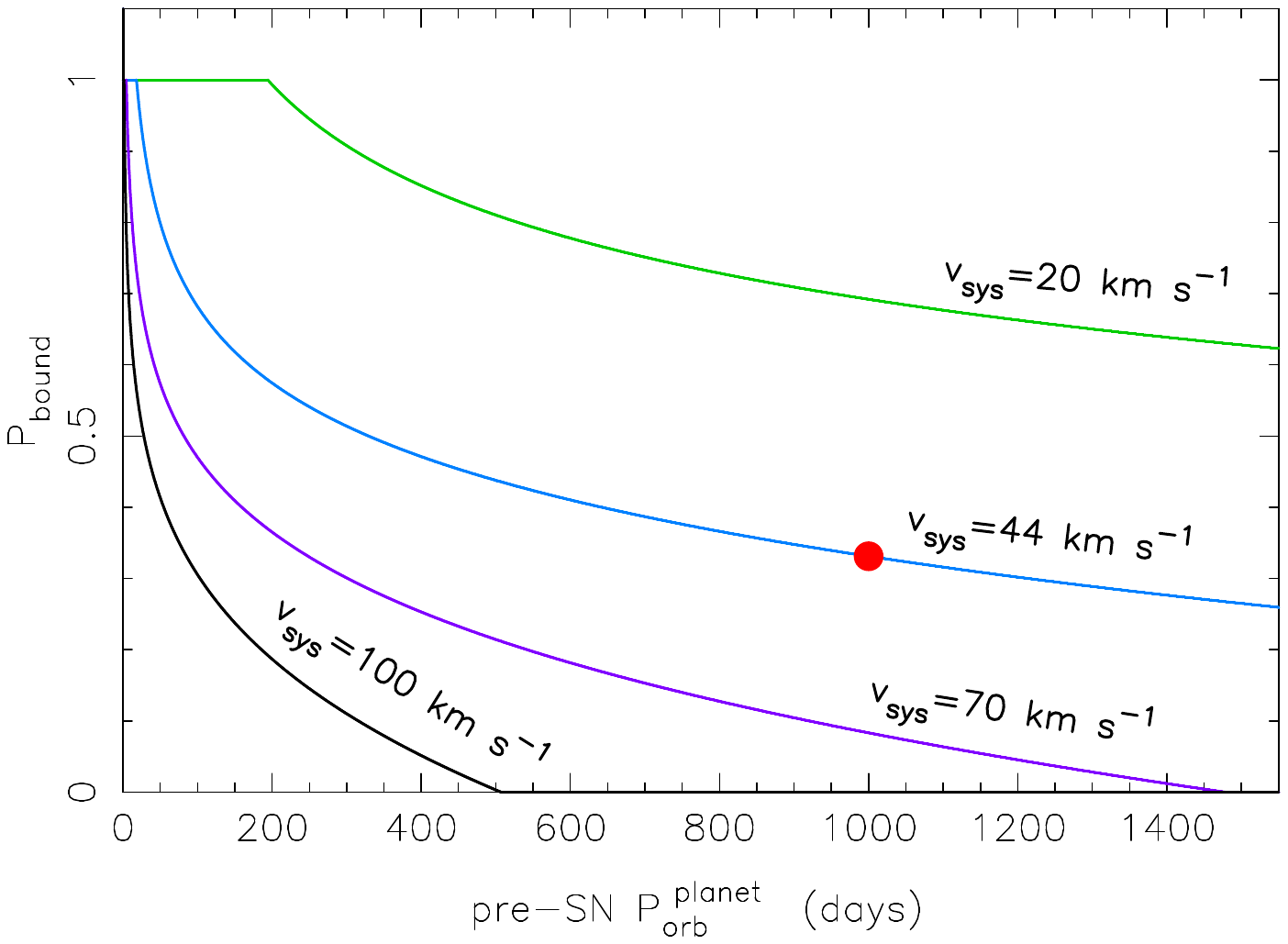}
\vspace{-1.2cm}
\end{center}
\caption{Probability of a planet surviving the SN explosion as a function of its pre-SN orbital period and the present-day systemic velocity of J0337. The red circle marks our default value with $v_{\rm sys}=44\;{\rm km\,s}^{-1}$ and $P_{\rm orb}^{\;\rm planet}=1000\;{\rm days}$, yielding $P_{\rm bound}=33\%$ (see text).
\label{fig:SN_survival}}
\end{figure}

Figure~\ref{fig:SN_survival} displays the probability of a planet surviving the SN explosion as a function of its pre-SN orbital period and the observed present-day systemic velocity of J0337. This probability can be calculated directly from the equations in \citet{hil83}: 
\begin{equation}
  P_{\rm bound}=\frac{1}{2}\, \bigg\{1+\left[\frac{1-2\Delta M/M -(v_{\rm sys}/v_{\rm rel})^2}{2\,(v_{\rm sys}/v_{\rm rel})}\right]\bigg\} \;,
\label{eq:bound}
\end{equation}
where $M$ is the total mass of the system, $\Delta M$ is the amount of material ejected in the SN, $v_{\rm rel}=\sqrt{GM/a}$ is the relative velocity between the triple system (``body~1'') and the planet (``body~2''), where $a$ is the separation between the centre of mass of the triple system and the planet (assumed to be in a pre-SN circular orbit), and finally where $v_{\rm sys}$ is the present-day observed velocity of J0337 with respect to its local standard of rest. The calculation is made by considering the triple system and the planet as an in-effect two-body problem, which is a good approximation given the large orbital separation of the planet and the fast-moving SN ejecta. In this picture, we therefore consider $v_{\rm sys}$ as the ``kick'' imparted on the triple system (``body~1''). For a general discussion of dynamical effects of asymmetric SNe in hierarchical multiple star systems, see \citet{pcp12}, and references therein. 

The parameter values needed for Eq.~(\ref{eq:bound}) are as follows at stage~4: $M=4.10\;M_\odot$ ($1.70 + 1.10 + 1.30\;M_\odot$), $\Delta M = 0.42\;M_\odot$, $v_{\rm sys}=44\;{\rm km\,s}^{-1}$ and $P_{\rm orb}^{\;\rm planet}=1000\;{\rm days}$. The latter two values are our default values: $v_{\rm sys}=44\;{\rm km\,s}^{-1}$ corresponds to our estimated value of the peculiar, that is, with respect to the local standard of rest, velocity of J0337 (see Sect.~\ref{subsec:gal_motion}) and $P_{\rm orb}^{\;\rm planet}=1000\;{\rm days}$ is selected as a typical value from our range of solutions to the SN event based on Monte Carlo simulations with $2\times 10^6$ trials, following the recipe of \citet{tkf+17}. 
In these Monte Carlo simulations, we have imposed a criterion of a post-SN orbital period of the planet of 910~days ($\pm 3\%$). This value is found by calculating backwards from stage~9 (present day) to stage~5 (right after the SN explosion, and before the formation of the two WDs), considering the mass lost from the system when the main-sequence star progenitors of the WDs evolve to fill their Roche lobes and undergo mass transfer \citep{tv23}. The orbital separation ratio (before and after the mass loss) will simply scale inversely to the total mass ratio: $a/a_0=M_0/M$. 
Based on our Monte Carlo simulations, we find maximum pre-SN orbital periods of \{2430; 1890; 910; 410\}~days 
for $v_{\rm sys}=$\{20; 44; 70; 100\}~${\rm km\,s}^{-1}$, respectively. The minimum pre-SN orbital period is about 100~days. It is limited by the condition of long-term dynamical stability of the system.

In terms of the effect of the interaction of the SN ejecta on the planet, the effect is expected to be negligible \citep{wlm75,ltr+15}. This is mainly due to the large orbital separation of the planet at the moment of the SN. In addition, the surface area of the planet is presumably quite small, and its mass density is likely to be larger than that of a main-sequence star ($\sim 1\;{\rm g\,cm}^{-3}$) or a gaseous planet ($\sim 0.1\;{\rm g\,cm}^{-3}$). For comparison the Moon has a mean mass density of $\sim 3.3\;{\rm g\,cm}^{-3}$.

The large orbital inclination of $\delta i_{\rm\Pi} = 119^{+16}_{-42}{}^\circ$ of the planet could have an origin in the SN explosion. It is certain that if the kick (or resulting recoil of the inner system) has a velocity vector that is not exactly in the orbital plane, then the system will be tilted. Whether or not this is sufficient to explain the orbital inclination of the planet is uncertain, and this inclination could also be due to dynamical interactions within the system. 

To summarise, we find from calculations in Fig.~\ref{fig:SN_survival} a most likely expected probability of the planet to survive the SN explosion of order 20\%--50\% (33\% for our default value). Therefore, {\em if} a planet is present in the J0337 system with an orbital period of about 3310~days, and {\em if} this planet formed from ejected CE material in stage~2, then we find it quite likely that the planet could have survived the kinematic impact of the SN that created the neutron star (NS) in this system. 

\subsubsection{Constraining the NS kick}
To place constraints on the kick velocity imparted directly on the NS at its formation is also possible, albeit rather complicated, and a proper treatment is beyond the scope of this paper. Using a step-by-step Monte-Carlo approach, however, where the kinematic effects are evaluated as in-effect two-body problems (step~1 consisting of ``body~1'': NS+WD$_{\rm i}$+WD$_{\rm o}$ and ``body~2'': planet; step~2 consisting of ``body~1'': NS+WD$_{\rm i}$ and ``body~2'': WD$_{\rm o}$; step~3 consisting of ``body~1'': NS and ``body~2'': WD$_{\rm i}$) we find a rough constraint on the NS kick of $w\simeq 110-125\;{\rm km\,s}^{-1}$. Here we made the assumption that the direction of recoil (or ``kick'') velocities is always isotropic. This is not likely to be the case because of the momentum in the pre-SN orbital plane in each step (only for the actual kick imparted onto the NS in step~3 is this a good assumption). Nevertheless, it seems to us that the resulting value of $w$ may not change by much taking such effects into account. A NS kick value of $w\simeq 110-125\;{\rm km\,s}^{-1}$ is somewhat on the low side for measured pulsar kicks \citep[with average values above $400\;{\rm km\,s}^{-1}$,][]{hllk05}. However, it is not an unusual small kick. Moreover, it is an obvious selection effect that the J0337 system exists, and thus the kick could not have been very large \citep[see also discussions in][]{tauris_formation_2014}.

In addition the small kick inferred for J0337, and its relatively small velocity relative to its local galactic co-rotating frame (see Fig.~\ref{fig:GalMotion}) are very similar to those observed for PSR~J1903+0327 \citep[][see especially their Fig. 10]{Freire_2011}, a system that is thought to have originated as a triple, which later became unstable because of the widening of the inner binary (then a low-mass X-ray binary) caused by the mass transfer to the NS \cite{Freire_2011,Portegies-Zwart_2011}. This, again, emphasizes the importance of small SN kicks for the preservation of hierarchical systems

\section{Conclusions}
We have extended the work of \citet{voisin_improved_2020} by using a 700-day longer dataset of the Nançay timing data of PSR J0337+1715 and, most importantly, by attempting to model the pulsar's residual long-term signal, which has an amplitude of $\sim 4.4 \rm \mu s$ over $\sim 3000$ days. 
To this end, we have complemented the numerical timing model for the triple stellar system introduced in \citep{voisin_improved_2020} with either an achromatic red-noise component or a planet in a hierarchical orbit around the triple system. Both models have been implemented in the code \texttt{Nutimo} \citep{voisin_improved_2020}. We note that the NRT ToAs considered in this study were extracted from total intensity profiles. In future work, we will consider using ToAs extracted from polarimetric profiles with the Matrix Template Matching \citep{vanStraten2006} technique, which was shown by \citet{guillemot_improving_2023} to significantly improve the quality of the NRT timing by compensating for imperfect polarisation calibration. (See footnote \ref{fn:zenodo} for software and data release.)

We have shown in Sec. \ref{sec:resdetails} that with sufficient data, the Planet model can display a clear signature that distinguishes it from red noise. Though we have attempted to select the best model using the AIC and BIC statistical criteria, we acknowledge that the difference between the models is within the systematic uncertainties in our numerical optimisation procedure given our current dataset. Nonetheless, both the red-noise and planet hypothesis are able to produce residuals that appear consistent with white noise (Fig. \ref{fig:LS3model}).

The best red-noise model contains three Fourier harmonics, the amplitude of which decreases with a steep power-law index of $\sim 2.7$ (or $\sim 5.4$ in the power convention, see Sec.\ref{sec:rndiscussion}). The Planet model implies a Moon mass object in a hierarchical $\sim 3000$-day orbit around the triple system, a $\sim 0.25$ eccentricity, and a $\sim 119^\circ$ inclination with respect to the fundamental plane of the triple system. Triple system, DM, and astrometric parameters are compatible within uncertainties such that they are essentially model independent. The full characterisations of the system assuming both models is presented in Tabs. \ref{tab:fitresults} and \ref{tab:modelspecific}.

From a Bayesian point of view, one may wonder which model is a priori more probable. Planets around pulsars are rare \citep{kerr_limits_2015,behrens_nanograv_2020,nitu_search_2022}; however PSR J0337+1715 is itself a unique system with an unusual formation channel. Thus, it is possible that the probability of existence of a planet in such a system is not as small as in the general population. In Sec. \ref{sec:formation}, we have proposed a scenario derived from \citet{tauris_formation_2014} in which the planet forms from the material expelled during the first common envelope phase in the evolution of the triple system and survives the subsequent stages of evolution.  In Sec. \ref{sec:planetdiscussion}, we have shown the likely short-term stability of the inferred orbit and noted that the confidence interval on its inclination with respect to the triple system is centred on the value of the exterior von Zeipel-Lidov-Kozai resonance. This resonance is narrow, and the confidence interval still allows for a mismatch. However, if such a correspondence were confirmed, it seems unlikely that such a coincidence would occur. This leads us to conjecture that this particular orbit may be more stable, possibly making this planet the only remnant of a larger population formed out of the expelled material. As a by-product of this study, we estimate a rather low supernova kick of $\sim 110-125 \rm km/s$, not unlike the kick of PSR~J1903+0327, which is believed to have originated as a triple system as well \citep{Freire_2011}. This supports the idea that low kicks might be a condition of survival of such systems.

On the other hand, the achromatic red noise is a common phenomenon usually associated to processes intrinsic to the pulsar that are yet to be fully understood \citep{lyne_switched_2010, melatos_pulsar_2014}. However, the amplitude of the signal has been shown empirically to scale with quantities such as the spin-down age, spin-down rate, or power of the pulsar \citep{hobbs_analysis_2010,lyne_switched_2010, shannon_assessing_2010}. In Sec. \ref{sec:rndiscussion}, we have shown that, according to those scalings, the red-noise amplitude necessary to explain the observed signal is a clear outlier by a factor of five to ten. Indeed, PSR J0337+1715 lies in the bulk of the millisecond pulsar distribution where red noise should be barely detectable with our data, if at all. A possible signature of strong red noise due to magnetospheric activity is its correlation to pulse profile variations, as observed in other pulsars \citep{lyne_switched_2010}, but we did not find any evidence of such variations here. Contrary to the planet hypothesis, we currently do not see any reason to speculate that such a discrepancy could be related to the particular formation channel of the system.

We have updated our previous limit on violations of the strong equivalence principle \citep{voisin_improved_2020}. In Sec. \ref{sec:SEP}, we have shown that the limit is somewhat model dependent, with $|\Delta|  <  2.29\times 10^{-6}$ assuming the best red-noise model and $|\Delta|  <  1.46\times 10^{-6}$ assuming the Planet model (both delimiting the 95\% confidence region). If the latter represents a 30\% improvement compared to \citet{voisin_improved_2020}, the former is 10\% worse. This systematic uncertainty emphasises the importance of selecting the correct model with future data.

\begin{acknowledgements}
This work was granted access to the HPC resources of MesoPSL financed
by the Region Ile de France and the project Equip@Meso (reference
ANR-10-EQPX-29-01) of the programme Investissements d’Avenir supervised
by the Agence Nationale pour la Recherche. 

\\
This research has made use of data obtained from or tools provided by the portal exoplanet.eu of The Extrasolar Planets Encyclopaedia.
\\
This work made use of the Scipy libraries (\url{www.scipy.org}).
\\
The Nan\c{c}ay Radio Observatory is operated by the Paris Observatory, associated with the French Centre National de la Recherche Scientifique (CNRS). We acknowledge financial support from the “Programme National Gravitation, Références, Astronomie, Métrologie (PNGRAM) funded by CNRS/INSU and CNES, France.
\\
We would like to thank the referee, Dr. Scott Ransom, for his insightful review, which allowed us to further improve this work.
\end{acknowledgements}

\bibliographystyle{aa} % style aa.bst
\bibliography{J0337} % your references Yourfile.bib

\begin{appendix}

\section{Second-order Keplerian R\oe mer delay} \label{apsec:roemerecc2}
The R\oe mer delay due to a pulsar orbiting in a binary system is given by \citep{blandford_arrival-time_1976},
\begin{equation}
\label{apeq:roemer}
\Delta_{\rm R} = x \left(\sin\omega\left(\cos E - e\right) + \sqrt{1-e^2}\cos\omega \sin E\right),
\end{equation}
where $x$ is the projected semi-major axis , $e$ is the orbital eccentricity, $\omega$ the argument of periastron, and $E$ the eccentric anomaly. 
The latter is solution of Kepler's equation, 
\begin{equation}
E - e\sin E = \sigma,
\label{apeq:kepl}
\end{equation}
where $\sigma = n (t-T_{\rm p})$ with time $t$, time of passage at periastron $T_{\rm p}$ and $n=2\pi/P$ for orbital period $P$.

Kepler's equation can be solved iteratively at any order in $e$ using a fixed-point method following $E_{n+1}  = \sigma + e\sin E_{n}$. 
Injecting $E_2$ into Eq. \ref{apeq:roemer} and Taylor expanding, we obtained the second-order R\oe mer delay,
\begin{eqnarray}
\label{apeq:roemerscnd}
\Delta_{\rm R} & = & x\left[\sin\phi \right. \\
& &      + \frac{e}{2}  \left(\cos\omega \sin 2\phi - \sin\omega \cos 2\phi - 3\sin\omega\right) \nonumber \\ 
& &  - \frac{e^2}{8}\left\{\left(\cos 2\omega + 4\right) \sin\phi - \sin 2\omega \cos\phi \right. \nonumber \\
& & \left.\left. - 3\cos 2\omega \sin 3\phi + 3 \sin 2\omega \cos 3\phi\right\} \right], \nonumber
\end{eqnarray}
where we define $\phi = n(t - T_{\rm asc})$ with $T_{\rm asc0} = T_{\rm p } - \omega/n$, which corresponds to the time of ascending node at order $e^0$. The first two lines correspond to the small-eccentricity timing model of \citet{lange_precision_2001} where one can identify the Laplace-Lagrange parameters $e\sin\omega$ and $e\cos\omega$. In all rigour we left here the term in $\sim e\cos\omega$, which is dismissed as an unnecessary additional constant in \citet{lange_precision_2001}  but can play a role in case of periastron precession \citep{susobhanan_exploring_2018, voisin_spider_2020}. Apart from this term, Eq. \eqref{apeq:roemerscnd} was also derived in \citet{Zhu_testing_2019}. 

We can see in Eq. \eqref{apeq:roemerscnd} that the effect of taking into account second-order terms is i) to modify the coefficients in front of first-harmonic terms and ii) to add signal at the third harmonic frequency. Without the third-harmonic contribution, i) would merely amount to an effective eccentricity and argument of periastron. However thanks to third harmonics this degeneracy is lifted and first-harmonic amplitudes become second-degree polynomials in eccentricity. 

Most importantly in the frame of this work, we can see that once limited to first order, this timing model is equivalent to the PL2 model. Moreover, we see that all five elements $x, e, \omega, T_p, n$ can be measured independently at this order up to an error $\bigcirc\left(e^2\right)$. The equivalence between PL2 as given by Eq. \eqref{eq:RN}, and this model approximately goes as follows: 
\begin{eqnarray}
\label{apeq:equivn}
n & = & 2\pi \nu, \\
x & = & A, \\
T_{\rm p} & = & T_{\rm ref} - \frac{\phi_2 - \phi_1}{\nu}, \\
\omega & = & 2\phi_1 - \phi_2, \\
\label{apeq:equive}
e & = & 2^{1-\gamma}. 
\end{eqnarray}
The approximation lies in the fact that we neglected the difference between emission time and arrival time, $\Delta = t_a-t_e$ Einstein delay excluded, such that Eqs. \eqref{apeq:equivn}-\eqref{apeq:equive} are valid up to corrections of order $\bigcirc\left(\Delta/P\right) <<1$. 

All subsequent harmonics are fully determined by these same five parameters, and therefore the Keplerian model has a higher predictive power than models such as PL3, which requires additional phase parameters. In addition it can be seen that contrary to the PLn models, amplitudes are not determined by a power law past PL2.

%-----------------------------------------------------------------------

\section{Simplified perturbative model \label{sec:pertubative}}

In order to assess how well the orbital motion of a small planet in a hierarchical orbit can be characterised we derive in this section an analytical model for a simplified configuration. The aim is to compute the leading order perturbation due to the planet beyond the Keplerian motion it induces on the barycentre of the triple stellar system (see below). 

We model the system by three bodies: the first one corresponds to the inner binary, the second one to the outer white dwarf and the last one to the planet. The extent of the inner binary is thus neglected. We also assume Newtonian dynamics. We note that this three-body modelling could also be applied to the triple system itself since it is also hierarchical. 

We denote $M_{0,1,2}$ the masses and $\vec{R}_{0,1,2}, \vec{V}_{0,1,2}$ the position and velocity vectors. Indexes refer to the three bodies in the same order as above. The inner binary is represented by the position and velocity of its barycentre $(\vR_0,\vV_0)$ and the sum of the masses of its two components. The justification for neglecting the dynamics of the inner binary is that due to its small extent relative to the separation with the planet it is much less sensitive to the planet perturbation than the outer binary.  
The problem is treated using Jacobi coordinates \citep[e.g.][]{murray_solar_1999},

\begin{eqnarray}
\label{eq:jac0}
 \vr_0 & = & \vB_{2}, \\
\label{eq:jac1}
 \vr_i & = & \vR_i - \vB_{i-1} \;\;(i>0),
\end{eqnarray}
where the centre of mass of the first $i+1$ bodies is
\begin{equation}
    \vB_i = \frac{1}{\sigma_i} \sum_{k=0}^i M_k \vR_k,
\end{equation}
with  $\sigma_i = \sum_{k=0}^i M_k$.

With these coordinates $\vr_1$ is the vector connecting the inner binary to the outer white dwarf, which we call the outer binary, and $\vr_2$ the vector going from the centre of mass of the triple stellar system to the planet. Given that the planet follows an orbit about five times wider than the size of the triple system and that its mass is negligible compared to those of the stars, we introduce the following scalings:
\begin{eqnarray}
 \epsilon &\sim& M_2/M_{0,1}  \ll 1, \\
 \eta & \sim & r_2/r_1 \sim R_2 / R_{0,1} \ll 1,
\end{eqnarray}
where we use the notation $V = \|\vec{V}\|$.

Using canonical coordinates and their conjugate momenta $\vP_{0,1,2}$, the Hamiltonian of the system is given by
\begin{equation}
\label{eq:H}
    H = \frac{1}{2}\sum_{k=0}^2 \frac{\vP_k^2}{M_k} - \sum_{0\leq k < l \leq 2}\frac{GM_kM_l}{\|\vR_l - \vR_k\|},
\end{equation}
where $G$ is the gravitational constant.

We now decompose the Hamiltonian in an dominant integrable part $H_0$ and a perturbing term $H_1$ using Jacobi coordinates Eqs.\eqref{eq:jac0}-\eqref{eq:jac1} and their conjugate momenta $\vp_i$. The kinetic part of the Hamiltonian Eq.\eqref{eq:H} keeps the same form after the substitution $\vP_k \rightarrow \vp_k, M_k \rightarrow m_k$ where $m_0 = \sigma_2$ (the total mass of the system) and $m_{i>0} = m_i \sigma_{i-1}/\sigma_i$. Noting that $\vp_0$ is the total momentum of the system, we choose it to be zero without loss of generality.
Further, the interaction term between zero and one is also left unchanged since $\vr_1 = \vR_1 - \vR_0$. On the other hand, the two terms involving the planet are expanded in powers of $\eta$. Collecting the terms,
\begin{eqnarray}
\label{eq:H0}
 H_0 & = &  \frac{\vp_1^2}{2m_1} + \frac{\vp_2^2}{2m_2} - \frac{Gm_1\sigma_1}{r_1} - \frac{Gm_2\sigma_2}{r_2}, \\
\label{eq:H1}
 H_1 & = &  - \frac{1}{2}\frac{\sigma_2}{\sigma_1}\frac{Gm_1m_2}{r_2}\left[\frac{3}{2}\left(\frac{\vr_1\cdot\vr_2}{r_2^2}\right)^2 - \left(\frac{r_1}{r_2}\right)^2\right]  + \bigcirc{\left(\eta^3\right)}.
\end{eqnarray}
One can see that $H_1$ is of order $\eta^2$ compared to the last term of Eq. \eqref{eq:H0} and of order $\eta\epsilon$ compared to the penultimate term, which justifies it being treated as a perturbation. We also note that $\sigma_2/\sigma_1 = 1+ \bigcirc{(\epsilon)}$, and therefore it is omitted in the following (however it would have to be kept if this methodology was applied to the triple stellar system).

The dominant Hamiltonian $H_0$ is recognised as the Hamiltonian of a mass $m_1$ orbiting around a central mass $\sigma_1$ and a mass $m_2$ independently orbiting around a central mass $\sigma_2$. Thus the solution is exactly given by two Keplerian orbits.

\subsection{Timing model corresponding to the dominant Keplerian Hamiltonian}
We consider only the largest delay, that is the R\oe mer delay,
\begin{equation}
\label{eq:dR}
    \Delta_R = -\frac{\nodot\cdot \vR_0}{c},
\end{equation}
where $\nodot$ is a unit vector along the direction going from the Solar System barycentre to the pulsar system barycentre and $c$ is the speed of light. We note that the motion of the pulsar is approximated here again to the motion of the barycentre of the inner binary $\vR_0$. A complete treatment could formally be obtained by replacing $\vR_0$ by $\vR_p = \vR_0 + \delta \vR_p$ where the second term accounts for the inner binary motion relative to its barycentre as well as every effect due to mutual interactions not accounted for in our treatment. 

In terms of Jacobi coordinates, 
\begin{equation}
\label{eq:R0}
    \vR_0 = -\frac{M_1}{\sigma_1} \vr_1 - \frac{M_2}{\sigma_2}\vr_2,
\end{equation}
where according to the exact Hamiltonian $H_0$ $\vr_{1,2}$ follow Keplerian orbits of the form 
\begin{equation}
\label{eq:vr}
    \vr_i = a_i\left(\begin{matrix}
    \cos E_i - e_i \\
    \sqrt{1-e_i^2} \sin{E_i} \\
    0
    \end{matrix}\right)_{F_i},
\end{equation}
where $F_i$ is the frame adapted to the orbital motion of coordinate $i$, that is with the direction of periastron along the $x$ axis and orbital angular momentum along the $z$ axis, $a_i$ is the semi-major axis, $E_i$ is the eccentric anomaly and $e_i$ the eccentricity. Kepler's equation relates $E_i$ to the mean anomaly $\mathcal{M}_i = 2\pi (t - T_i)/P_i$, where $P_i$ is the orbital period and $T_i$ the time of periastron passage,
\begin{equation}
    E_i - e_i \sin E_i = \mathcal{M}_i.
\end{equation}

In order to express the Keplerian R\oe mer delay one needs to express both $\nodot$ and $\vR_0$ in the same frame. We choose a frame where $\nodot$ is the third basis vector. We call it the observer's frame $F_O$ (who would be standing at the Solar System barycentre). Orbital inclinations $i_j$ are then defined relative to $\nodot$ and longitudes of ascending nodes $\Omega_j$ define a rotation around $\nodot$. Together with the arguments of periastron $\omega_j$ these angles define the rotations relating the observer's frame and $F_j$,
\begin{equation}
    \vr_j\lvert_{F_O} = R_{\Omega_j} R_{i_j} R_{\omega_j} \vr_j\lvert_{F_j},
\end{equation}
where $ R_{\Omega_j}, R_{i_j}, R_{\omega_j}$ are rotation matrices (see e.g. \citet{beutler_methods_2004}), and $\vr_j\lvert_{F}$ refers to the coordinate of $\vr_j$ in frame $F$. From there, the formula expressing the Keplerian R\oe mer delay $\Delta_j = -\nodot\cdot \vr_j/c$ is well known \citep[e.g.][]{lyne_pulsar_2012, hobbs_tempo2_2006},
\begin{eqnarray}
    \label{eq:dRk}
    \Delta_j & = & -\frac{a_j\sin i_j}{c} \\
    & & \left(\sqrt{1-e_j^2}\sin E_j\cos\omega_j + \cos E_j\sin\omega_j  - e_j\sin\omega_j\right), \nonumber
\end{eqnarray}
We note that the last term of Eq. \eqref{eq:dRk} is usually omitted because it contributes only a constant delay to the timing solution, but we write it here for completeness as it can become important when orbits are perturbed \citep[e.g.][]{voisin_spider_2020}. Let us also remark that this formula is independent of $\Omega_j$ and that $a_j$ cannot be separated from $\sin i_j$, which leads to the impossibility of independently measuring these three parameters with the Keplerian R\oe mer delay alone. 
Inserting Eq.\eqref{eq:R0} into \eqref{eq:dR}, we can write 
\begin{equation}
\label{eq:dR0}
    \Delta_R = -\sum_{j = 1}^2\frac{M_j}{\sigma_j} \Delta_j.
\end{equation}

\subsection{Perturbation}
We now turn to the perturbative treatment of $H_1$, Eq.\eqref{eq:H1}, using Laplace-Lagrange's variation of the orbital elements. In order to retain only the strongest effects and to keep the treatment as simple as possible we neglect all eccentricities, and we retain only secular terms. We note that we neglect eccentricity only in perturbative terms, not in leading order terms, that is Eq.\eqref{eq:dRk}. Thus we seek a R\oe mer delay of the form Eq.\eqref{eq:dR0} with
\begin{equation}
\label{eq:roepert}
    \Delta_j = \Delta_j^{(0)} + \delta\Delta_j,
\end{equation}
where $\Delta_j^{(0)}$ is given by Eq.\eqref{eq:dRk}, and 
\begin{equation}
\label{eq:deltaDelta}
    \delta\Delta_j = \delta\left(-\frac{a_j\sin i_j}{c}\left(\sin \left(n_j t + \phi_j\right)\right) \right),
\end{equation}
which denotes the variation of the leading order term in eccentricity of Eq.\eqref{eq:dRk}. For convenience we introduced $\phi_j = \omega_j -n_jT_j$ such that at leading order $n_j t + \phi_j = E_j + \omega_j$. 

The outer-binary term in Eq.\eqref{eq:dR0} is dominant since $r_2 M_2/\sigma_2 \sim (\epsilon/\eta) r_1 M_1/\sigma_1 $ and in the present case $ \epsilon \ll \eta$. Therefore we focus on the perturbation of $\Delta_1$. Without loss of generality we choose a frame such that $\Omega_2 = 0$ such that $\Omega_1$ is defined relative to the longitude of ascending node of the planet. 

In order to get rid of periodic terms we average $H_1$ over $\mathcal{M}_1$, being understood that $P_1 \gg P_2$, and obtain $\bar H_1 = (2\pi)^{-1}\int_0^{2\pi} H_1 {\rm d}\mathcal{M}_1$. Although straightforward, this operation is very lengthy. We used the formal calculus software Sage \citep[][]{developers_sagemathsage_2022} in order to manipulate the large expressions generated, without any particular difficulty. Therefore in what follows, we only report the important steps and the main results. 
%The corresponding Sage notebook is available as an online supplementary content to this paper. 

The resulting perturbing Hamiltonian does not depend on $\omega_{1,2}, T_1, E_1, e_{1,2}, \Omega_2$. The relevant Laplace-Lagrange variational equations are
\begin{eqnarray}
\label{eq:LL1}
\dot i_1 & = &  -\frac{1}{n_1 a_1^2\sin i_1} \pderiv{H_1}{\Omega_1}, \\
\dot \Omega_1 & = & \frac{1}{n_1 a_1^2\sin i_1} \pderiv{H_1}{i_1}, \\
\dot \phi_1 & = & -\frac{1}{n_1a_1}\left(\frac{1}{a_1\tan i_1}\pderiv{H_1}{i_1} + 2\pderiv{H_1}{a_1}\right),\\
\label{eq:LL4}
\dot a_1 &=& \dot n_1 = \dot e_1 =0.
\end{eqnarray}
We seek solutions of the form $x(t) = x^{(0)} + \delta x(t)$ where $x^{(0)}\in\{a_1, n_1, i_1, \Omega_1, \phi_1\}$ are the constant orbital elements that are solution of the unperturbed system. To leading order in perturbation we may thus substitute $x \rightarrow x^{(0)}$ in the right-hand side of Eqs.\eqref{eq:LL1}-\eqref{eq:LL4} and integrate in order to obtain $\delta x =\int \dot x {\rm d} t$. In order to focus on the main contributions we keep only terms diverging with time, that is terms $\propto E_2$. In the following, we drop the $(0)$ exponent for readability since orbital elements always refer to the unperturbed solution, that is $x=x^{(0)}$ unless otherwise stated. 

Substituting $x \rightarrow x + \delta x(t)$ into Eq. \eqref{eq:deltaDelta} we obtained an expression of the form 
\begin{equation}
\label{eq:delta1}
    \delta\Delta_1 = \alpha t \left(\gamma_c\cos E_1 + \gamma_s \sin E_1\right),
\end{equation}
where $E_1 = n_1(t-T_1)$ at leading order in eccentricity and 
\begin{eqnarray}
\label{eq:alpha}
\alpha & = &\frac{3}{8}\frac{Gm_2 a_1}{c n_1a_2^3},\\
\label{eq:gammas}
\gamma_s & =& -\sin\Omega_1\cos i_1 \sin i_2 \\
& & \left(\cos i_1 \cos i_2 + \cos\Omega_1 \sin i_1 \sin i_2\right), \nonumber\\
\label{eq:gammac}
\gamma_c & =& \sin i_1\left(\cos^2 i_1 \cos^2 i_2 - \frac{10}{3}\right) \\
 & & +\cos \Omega_1 \sin i_2 \cos i_2 \cos i_1 \left(1 + 2\sin^2i_1\right) \nonumber \\
 & & +(\cos \Omega_1 \sin i_2)^2 \sin i_1 \left( 1+\sin^2 i_1\right). \nonumber
\end{eqnarray}

In order to validate the ability of our perturbative approach to capture the main features of the R\oe mer delay induced by the planet we have fitted $\Delta_1$, obtained by substituting Eq.\eqref{eq:delta1} in Eq. \eqref{eq:roepert}, to the numerically computed R\oe mer signal. The fitting was done using a Levenberg-Marquardt method\footnote{Implemented in the Scipy least\_square routine.}. The result is shown in Fig.\ref{fig:comparison}, where the Keplerian component $\Delta_1^{(0)}$ is also reported, and we can see that the perturbative formula successfully captures the dominating frequencies, that is the planet and outer binary orbital frequencies. The main residual appears at the first harmonic of the planet orbital frequency. 

\begin{figure}
\begin{center}
    \includegraphics[width=0.5\textwidth]{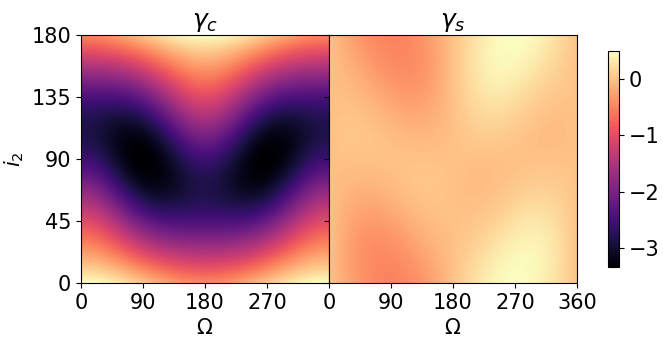}
\end{center}
\caption{Maps representing the values of $\gamma_c$ (left) and $\gamma_s$ (right) as a function of $(\Omega_1,i_2)$ for $i_1=39\deg$.\label{fig:gammamaps}}
\end{figure}

The parameters $\alpha\gamma_c, \alpha\gamma_s$ can in principle be fitted independently due to the Fourier decomposition theorem. Since $\alpha$ only occurs as a common scaling in Eq.\eqref{eq:delta1}, we focus on $\gamma_c, \gamma_s$. If their relation to $\Omega_1, i_2$ is not degenerate, Eqs. \eqref{eq:gammas}-\eqref{eq:gammac}, then this allows the degeneracies of the Keplerian component to be lifted and full characterisation of the planet's orbit and mass. In order to check for the absence of degeneracy, we plotted in Fig.\ref{fig:gammamaps} $\gamma_c(\Omega_1, i_2), \gamma_s(\Omega_1, i_2)$ for $i_1 =39\deg$, which approximately corresponds to the observed inclination of the outer binary. We see that although not degenerate, $\gamma_c$ is about five times larger than $\gamma_s$ for this particular inclination $i_1$. Given the overall weakness of the signal, it might therefore be difficult to constrain accurately both $i_1$ and $\omega_2$ in this particular case.

\FloatBarrier
\section{Best-fit systematics}

\label{ap:fitting}

Table \ref{aptab:modelcomp} summarises all the fits that have been performed, while Table \ref{tab:modelcomp} only contains the best fit for each model. The procedure involved using either the mean, `mean, 'or the maximum-posterior probability (including prior), `max' element of the MCMC sample of each model as the starting point for a deterministic fit using the \texttt{Minuit} library. Then, this `refit' converges to the local optimum. The fact that the local optima generally depend on their starting points is evidence for a rough likelihood surface with multiple nearby local maxima. We note that each refit has been iterated one more time in order to make sure that this difference was not related to an approximate convergence of the algorithm. 

Surprisingly, the PL4 model produced less optimal log-likelihood than the PL3 model, although the latter is nested in the former. In order to attempt to find a better solution we ran two MCMC: one starting from the last chain elements of the PL3 run and another from the PL5 run, which we call PL4bis. However, none of these fits performed better than PL3, suggesting additional complications in the probability landscape.

The Planet model behaves somewhat differently than the others. Indeed it is the one where the likelihood difference between the mean and max solutions is by far the largest, and the best fit is given by the max, suggesting that the posterior probability distribution is asymmetric or has multiple modes. This is consistent with its more irregular correlation plots, Fig. \ref{fig:subcornerplanet}, compared to the PL3 correlations in Fig. \ref{fig:subcornerpl3}. We also note that although the Planet max solution has the best log-likelihood of all models, the gain after refitting is more moderate than in other models.

To sum up, Table \ref{aptab:modelcomp} shows that a difference of a few units in log-likelihood may not be safely considered as significant given the systematic uncertainty of the fitting procedure. 

\begin{table*}%[h]
        \centering
        \caption{All fits carried out for each model assuming GR. \label{aptab:modelcomp}}
        \begin{tabular}{lllllll}
                Model & Npar  &  $\chi^2$  & $R_{\chi^2}$  &     $\Delta$AIC        &            $\Delta$BIC &\\
                \hline
                PL2/Kepl1 & 30 & & & & \\
                %\hline
                \;\; {mean}    &    & { 15531.9 } & { 1.24814 } &  {15591.9}  & { 15814.8} \\
                \;\; max    &    &  8.09  &  0.00065  &  8.09  &  8.09  \\
                \;\; refit max    &    &  4.64  &  0.000373  &  4.64  &  4.64  \\
                \;\;{ refit mean}    &    &  {-0.915}  &  {-7.35e-05}  &  {-0.915}  &  {-0.915 } \\
                \hline
                PL3 & 31 & & & & \\
                %\hline
                \;\; {mean}    &    &  {15543.4 } &{  1.24917}  & { 15605.4}  &  {15835.8}  \\
                \;\; max    &       &  7.93  &  0.000637  &  7.93  &  7.93  \\
                \;\; refit max    &       &  -31.5  &  -0.00253  &  -31.5  &  -31.5  \\
                \;\; {refit mean}    &       &  {-30.7}  &  {-0.00247}  & { -30.7}  &  {-30.7}  \\
                \hline
                PL3DM10 & 39 & & & & \\
                \;\; {mean}  &    &  {15496.8}  &  {1.24622}  &  {15574.8}  &  {15864.6}  \\
                \;\; max  &    &  -0.509  &  -4.09e-05  &  -0.509  &  -0.509  \\
                \;\; refit max  &    &  -1.87  &  -0.000151  &  -1.87  &  -1.87  \\
                \;\; {refit mean}  &    &  {-2.22}  &  {-0.000179}  &  {-2.22}  &  {-2.22}  \\
                \hline
                PL4 & 32 & & & & \\
                %\hline
                \;\; {mean}    &      &  {15569.8}  &  {1.25139}  & { 15633.8 } &  {15871.6}  \\
                \;\;  max    &      &  7.56  &  0.000608  &  7.56  &  7.56  \\
                \;\; refit max   &      &  -50.3  &  -0.00404  &  -50.3  &  -50.3  \\
                \;\; {refit mean }   &      & { -53.2}  &  {-0.00427}  & { -53.2}  &  {-53.2 } \\
                \hline
                PL4 bis & 32 & & & & \\
                \;\; {mean}    &      &  {15558.3}  &  {1.25047 } & { 15622.3}  &  {15860.1}  \\
                \;\;  max    &      &  9.62  &  0.000773  &  9.62  &  9.62  \\
                \;\; refit mean   &      &  -38.2  &  -0.00307  &  -38.2  &  -38.2  \\
                \;\; refit max   &      &  -36  &  -0.00289  &  -36  &  -36  \\
                \hline
                PL5 & 33 & & & & \\
                \;\; {mean}    &    & { 15535.8}  & { 1.24876 } &{  15601.8 } & { 15847.1 } \\
                \;\; max    &    &  4.82  &  0.000388  &  4.82  &  4.82  \\
                \;\; refit max    &    &  -24.2  &  -0.00195  &  -24.2  &  -24.2  \\
                \;\; {refit mean}   &    &  {-29.3}  &  {-0.00236 } & { -29.3}  & { -29.3 } \\
                \hline 
                planet & 32 & & & & \\
                \;\; {max}   &    &  {15520.9}  &  {1.24746}  &  {15584.9}  &  {15822.7 } \\
                \;\; mean    &   & 520  &  0.0418  &  520  &  520   \\
                \;\; refit mean  &    &  -1.76  &  -0.000141  &  -1.76  &  -1.76  \\
                \;\; {refit max}  &    &  {-6.56}  &  {-0.000527}  &  {-6.56}  &  {-6.56}  \\
        \end{tabular}
    	\tablefoot{ The reduced $\chi^2$, $R_{\chi^2}$, is equal to $\chi^2/\rm Ndof$ where  $\mathrm{Ndof} = 13534 - \mathrm{Npar}$. AIC and BIC are the Akaike and Bayesian information criteria, respectively. For each model the first line is the reference fit subtracted from all the other lines. The last line is the best fit of each given model that appears in Table \ref{tab:modelcomp} (except for `PL4 bis'). The label `mean' indicates that it is the mean of the MCMC sample, `max' shows that it is the element in the sample with the highest likelihood, and `refit' means that a local fit using the \texttt{Minuit} library (see sec.\ref{sec:modcomp}) from either the `mean' or `max' solution has been performed. For model `PL4', `bis' refers to a separate MCMC run (see text).}
\end{table*}

%-----------------------------------------------------------------------------------------------------------

\FloatBarrier

\section{Posterior distribution functions of model-specific parameters \label{ap:correlplots}}

\begin{figure*}
        \begin{center}
                \includegraphics[width=1\textwidth]{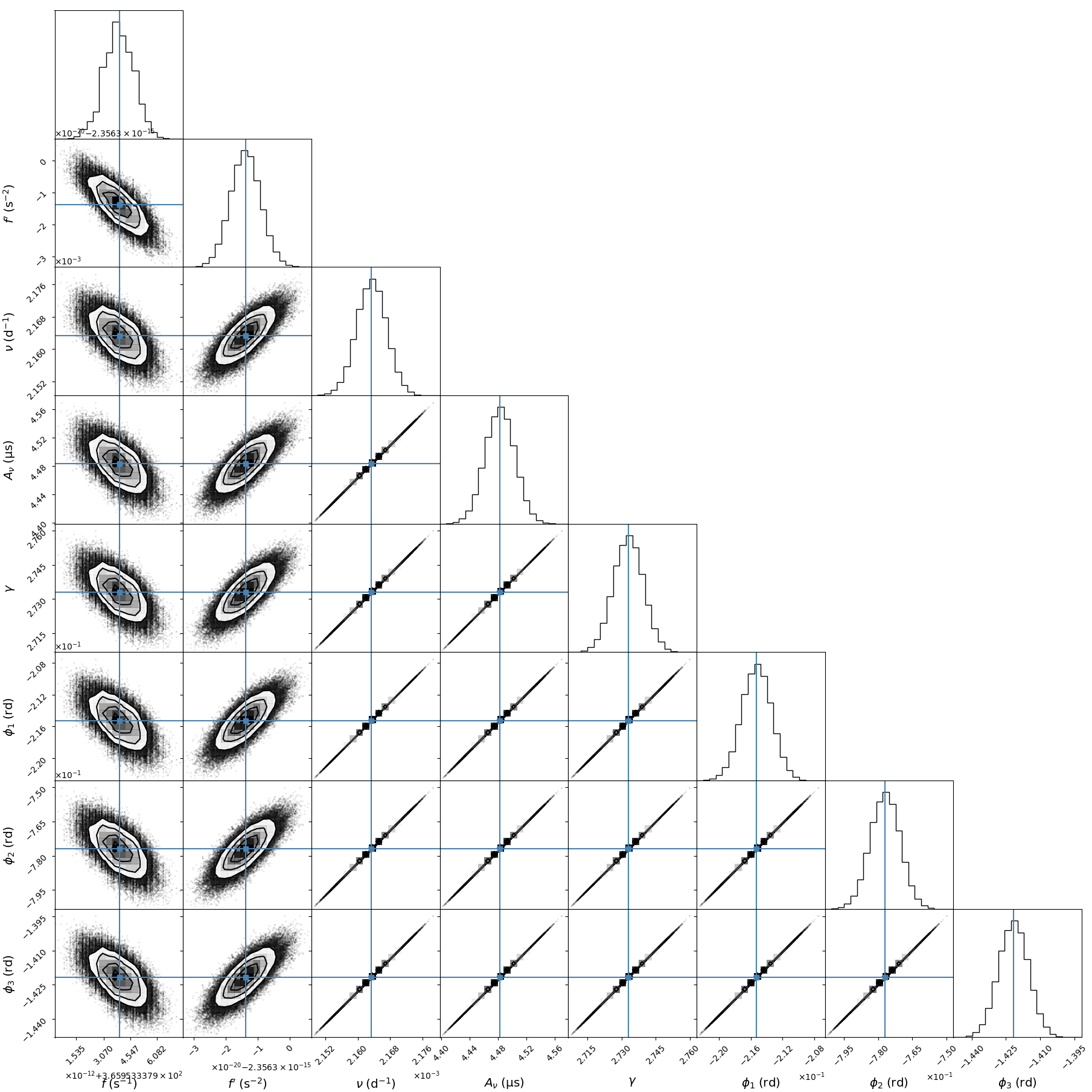}
        \end{center}
        \caption{Correlation plot of the posterior distribution function of the PL3 model using the corresponding MCMC-generated sample, and restricted to $\bar f, \bar \dot f$ as well as PL3 specific parameters.   \label{fig:subcornerpl3}}
\end{figure*}

\begin{figure*}
        \begin{center}
                \includegraphics[width=1\textwidth]{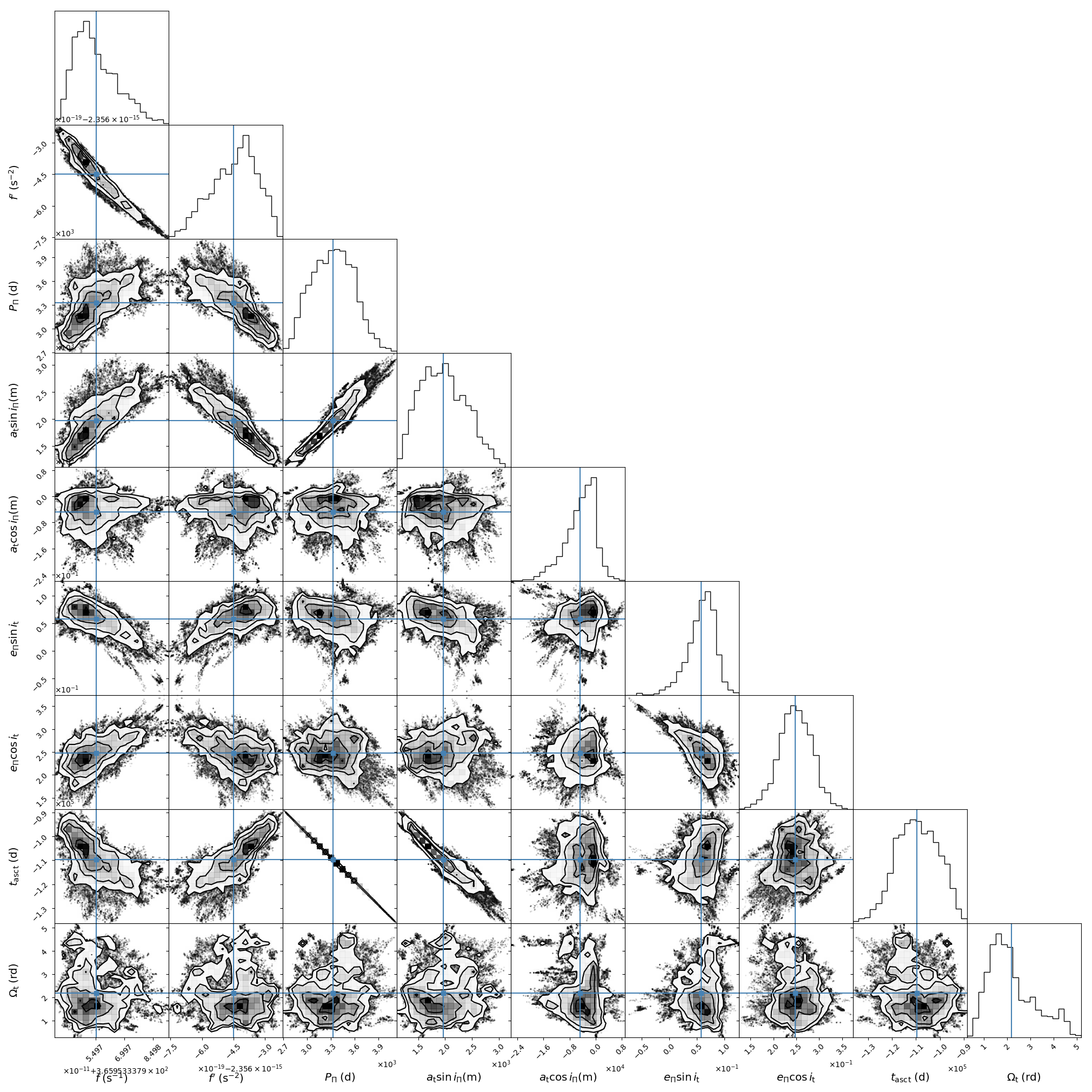}
        \end{center}
        \caption{Correlation plot of the posterior distribution function of the PlanetGR model using the corresponding MCMC-generated sample, and restricted to $\bar f, \bar \dot f$ as well as Planet specific parameters.   \label{fig:subcornerplanet}}
\end{figure*}

Model specific parameters, as reported in Table \ref{tab:modelspecific}, are not significantly correlated with other parameters of the timing model except for the re-scaled spin parameters $ \bar f, \bar \dot f$. Therefore we show in this appendix the corresponding `correlation plots' (or `corner plots') showing the posterior distribution function of the PL5 and Planet model marginalised along all but two dimensions (or one across the diagonal), Fig. \ref{fig:subcornerpl3} and Fig. \ref{fig:subcornerplanet} respectively. Both correspond to the version of these models that assumes GR (namely PL5GR and PlanetGR). Full correlation plots are provided as online supplementary material.

\onecolumn
\section{Additional material}
\FloatBarrier
\begin{figure*}
        \centering
        \begin{center}
                \includegraphics[width=15.5cm]{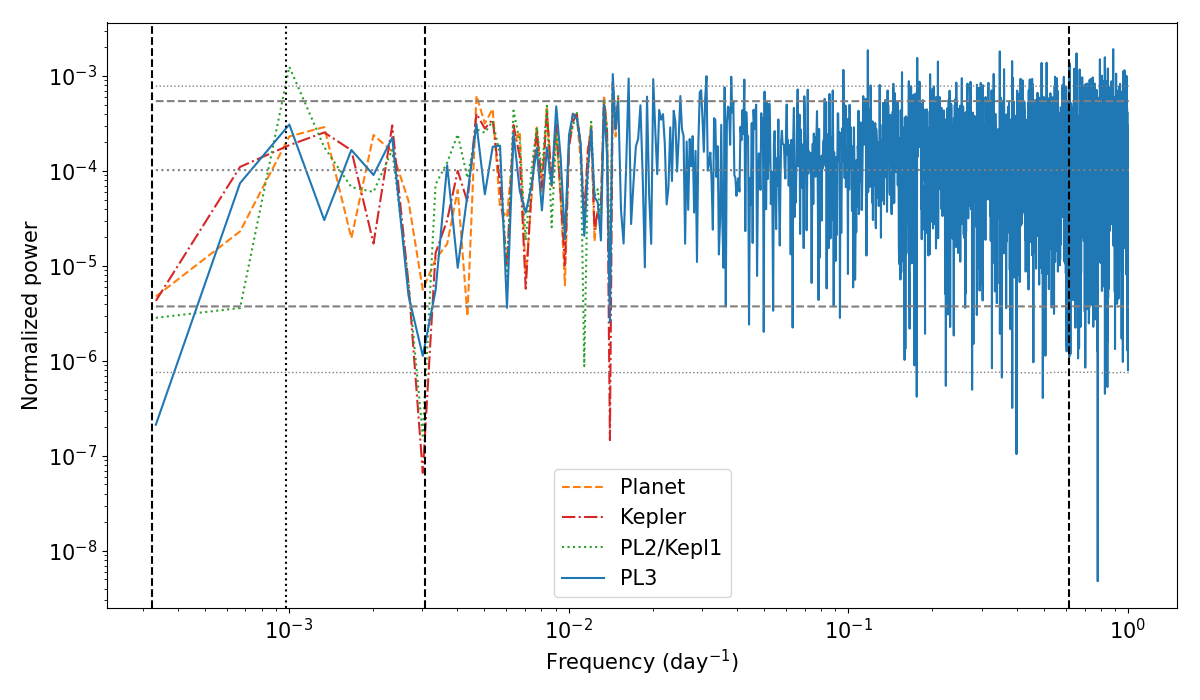}
        \end{center}
        \caption{Periodogram of the post-fit residuals, sampled at $1/T$ where $T$ is the time span of observations, and ranging in frequency from $1/T$ to $1\rm day^{-1}$. Vertical dashed lines represent the best-fit planet orbital frequency, outer binary and inner binary orbital frequencies, from left to right. The vertical dotted line marks the third harmonic of the planet orbital frequency. Horizontal grey lines describe the $95\%$ region obtained by bootstrapping white noise: the central dotted line is the median, dashed lines delineate the 2.5\% and 97.5\% levels, and thin dotted lines the 0.5\% to 99.5\% interval. For clarity, these lines have been smoothed as they are otherwise noisy. In the low-frequency part of the plot, until $5/P_{\rm O}\simeq 1.5\times 10^{-2} \rm day^{-1}$ where $P_{\rm O}$ is the outer binary period, we show the periodograms of the planet (dashed orange), Kepler (dash-dotted red), PL2/Kepl1 (dotted green), and PL3 (solid blue) models. In the higher frequency part only PL3 is shown for clarity as the difference with other models becomes much smaller. \label{fig:LS3model}}
\end{figure*}

\FloatBarrier
\begin{table*}
        \caption{\label{tab:fitresults} Mean values of the MCMC runs of the PL3 and Planet models with their $68\%$ median confidence intervals. }
        \centering
        \renewcommand{\arraystretch}{1.15}
        \begin{footnotesize}
                \begin{tabular}{lcll}
                        Parameter & Symbol & PL3GR & PlanetGR
                        % Value \MK{MORE READABLE TO MOVE UNITS TO LEFT COLUMN}: PF- this has been addressed. Table now is in a format more similar to current ways of displaying timing parameters in the literature
                        \\
                        \hline
                        \multicolumn{4}{c}{Fixed values} \\
                        \hline
                        Reference epoch ($\mathrm{MJD}$) & $T_{\mathrm{ref}}$ & 56492 &  \\
                        Position epoch ($\mathrm{MJD}$) & $T_{\mathrm{pos}}$ & 57205 & \\
                        %Data interval ($\mathrm{MJD}$) & & $56492 - 59480$ & \\
                        %Solar system ephemeris & & DE430 & \\
                        %        Start of data ($\mathrm{MJD}$) &  & XXXX  \\
                        %       End of data ($\mathrm{MJD}$) &  & XXXX  \\
                        %      Number of TOAs  & & 9303 \\
                        %    Residual root mean square ($\mu$s) & & 1.8 \\
                        %                Reduced $\chi^2$ & & 1.05 \\
                        \hline
                        \multicolumn{4}{c}{Fitted values} \\
                        \hline
                        Right ascension & $\alpha$   &   $3^{\rm h}37^{\rm m}43^{\rm s}.82700(43)_{-67}^{+66}$   &   $3^{\rm h}37^{\rm m}43^{\rm s}.82700(94)_{-64}^{+63}$\\ 
                        Declination & $\delta$    &   $17^{\rm deg}15^{\rm m}14^{\rm s}.817(63)_{-37}^{+38}$   &   $17^{\rm deg}15^{\rm m}14^{\rm s}.817(75)_{-33}^{+40}$\\ 
                        Distance ($\mathrm{kpc}$) & $d$    &   $1160_{-51}^{+53} $   &   $1182_{-55}^{+53} $\\ 
                        Right-ascension proper motion ($\mathrm{mas \, yr^{-1}}$) & $\mu_{\rm \alpha}$    &   $6.1(37)_{-68}^{+70}$   &   $5.9(82)_{-80}^{+70}$\\ 
                        Declination  proper motion ($\mathrm{mas\, yr^{-1}}$) & $\mu_{\rm \delta}$     &   $-5.(34)_{-26}^{+25}$   &   $-4.(83)_{-25}^{+29}$\\ 
                        Radial proper motion ($\mathrm{mas\, yr^{-1}}$) & $\mu_{\rm d}$     &   $5.(63)_{-32}^{+31}$   &   $5.(52)_{-28}^{+32}$\\ 
                        Dispersion measure ($\mathrm{pc}\,\mathrm{cm}^{-3}$) & $\mathrm{DM}$   &   $21.316(46)_{-12}^{+12}$   &   $21.316(46)_{-11}^{+11}$\\ 
                        Dispersion measure variation ($\mathrm{pc}\,\mathrm{cm}^{-3}\,\mathrm{yr}^{-1}$) & $\mathrm{DM}'$   &   $(1.7)_{-2.1}^{+2.1} \times 10^{-5}$   &   $(1.6)_{-2.0}^{+1.9} \times 10^{-5}$\\ 
                        \hline
                        Re-scaled spin frequency ($\si{Hz}$) & $\bar{f}$    &   $365.953337902243(85)_{-81}^{+81}$   &   $365.9533379022(56)_{-11}^{+11}$\\ 
                        Re-scaled spin frequency derivative ($10^{-15}\, \si{Hz \, s^{-1}}$) & $\bar{f}'$   &   $-2.35631(40)_{-45}^{+48} \times 10^{-15}$   &   $-2.356(45)_{-12}^{+11} \times 10^{-15}$\\ 
                        \hline
                        \multicolumn{4}{l}{\textit{Inner orbit $\rm I$ of pulsar $\rm p$ w.r.t. inner-binary centre of mass $\rm b$}} \\
                        Orbital period ($\mathrm{d}$) & $P_{\rm I}$    &   $1.62939900(75)_{-42}^{+42}$   &   $1.62939900(83)_{-41}^{+38}$\\ 
                        Projected semi-major axis (lt-s) & $a_{\rm p}\sin i_{\rm I}$   &   $1.2175276(17)_{-56}^{+56}$   &   $1.2175276(25)_{-56}^{+54}$\\ 
                        Inclination offset ($^{\circ}$) & $\delta i = i_{\rm I} - i_{\rm O}$    &   $1.2_{-2.4}^{+2.3} \times 10^{-3}$   &   $8_{-23}^{+21} \times 10^{-4}$\\ 
                        Laplace-Lagrange & $e_{\rm I}\cos\omega_{\rm p}$    &   $6.937(38)_{-40}^{+40} \times 10^{-4}$   &   $6.937(32)_{-42}^{+37} \times 10^{-4}$\\ 
                        Laplace-Lagrange & $e_{\rm I}\sin \omega_{\rm p}$   &   $-8.59(37)_{-41}^{+40} \times 10^{-5}$   &   $-8.59(42)_{-35}^{+36} \times 10^{-5}$\\ 
                        Time of ascending node ($\mathrm{MJD}$) & ${t_{\mathrm{asc}}}_{\rm p}$   &   $55917.15894(93)_{-19}^{+19} $   &   $55917.15894(89)_{-17}^{+19} $\\ 
                        Long. of asc. nodes offset ($^\circ$) & $\delta\Omega = \Omega_{\rm b} - \Omega_{\rm p}$   &   $(1.7)_{-1.3}^{+1.3} \times 10^{-4}$   &   $(1.5)_{-1.2}^{+1.2} \times 10^{-4}$\\ 
                        \hline
                        \multicolumn{4}{l}{\textit{Outer orbit $\rm O$ of $\rm b$ w.r.t. triple-system centre of mass $\rm t$}} \\
                        Orbital period ($\mathrm{d}$) & $P_{\rm O}$   &   $327.25512(55)_{-10}^{+10}$   &   $327.25512(61)_{-11}^{+10} $\\ 
                        Projected semi-major axis (lt-s) & $a_{\rm b}\sin i_{\rm O}$   &   $74.672344(35)_{-23}^{+22}$   &   $74.672344(31)_{-22}^{+21}$\\ 
                        Co-projected semi-major axis (lt-s) & $a_{\rm b}\cos i_{\rm O}$    &   $91.4(10)_{-37}^{+37}$   &   $91.4(02)_{-32}^{+35}$\\ 
                        Laplace-Lagrange & $e_{\rm O}\cos\omega_{\rm b}$   &   $0.03511469(40)_{-27}^{+28} $   &   $0.03511469(43)_{-27}^{+28} $\\ 
                        Laplace-Lagrange & $e_{\rm O}\sin \omega_{\rm b}$   &   $-0.00352492(40)_{-18}^{+19} $   &   $-0.00352492(22)_{-18}^{+19} $\\ 
                        Time of ascending node ($\mathrm{MJD}$) & ${t_{\mathrm{asc}}}_{\rm b}$   &   $56230.195329(47)_{-82}^{+83} $   &   $56230.195329(01)_{-78}^{+79}$\\ 
                        Longitude of ascending node ($^{\circ}$) & $\Omega_{\rm b}$    &   $145.9_{-1.5}^{+1.4} $   &   $148.0_{-1.5}^{+1.7}$\\ 
                        \hline
                        Inner mass ratio & $m_{\rm i}/m_{\rm p}$   &   $0.1373(76)_{-12}^{+12}$   &   $0.1373(77)_{-12}^{+12}$\\ 
                        ToA uncertainty re-scaling & $\mathrm{EFAC}$   &   $1.11(77)_{-73}^{+76}$   &   $1.11(72)_{-79}^{+75}$\\ 
                        \hline
                        \multicolumn{4}{c}{Derived values} \\
                        \hline
                        Parallel proper motion ($\mathrm{km\, s^{-1}}$) & $V_{\parallel}$   &   $29.77_{-1.1}^{+0.90}$   &   $29.(72)_{-83}^{+76}$\\ 
                        Plane-of-sky proper motion ($\mathrm{km\, s^{-1}}$) & $V_{\bot}$   &   $43.0_{-2.0}^{+2.2}$   &   $41.4_{-2.2}^{+2.1}$\\ 
                        \hline
                        Spin frequency ($\si{Hz}$) & $f$   &   $365.9533629(82)_{-13}^{+13} $   &   $365.9533629(80)_{-12}^{+12} $\\ 
                        Spin frequency derivative ($10^{-15} \, \si{Hz \, s^{-1}}$) & $\dot f$   &   $-2.29(06)_{-40}^{+43} \times 10^{-15}$   &   $-2.29(67)_{-43}^{+41} \times 10^{-15}$\\ 
                        \hline
                        \multicolumn{4}{l}{\textit{Inner orbit $\rm I$ of $\rm p/b$}} \\
                        Semi-major axis (lt-s) & $a_{\rm p}$   &   $1.924(46)_{-48}^{+48}$   &   $1.924(39)_{-43}^{+46}$\\ 
                        Orbital inclination ($^{\circ}$) & $i_{\rm I}$   &   $39.2(47)_{-11}^{+11}$   &   $39.2(48)_{-11}^{+10}$\\ 
                        Orbital eccentricity & $e_{\rm I}$   &   $6.990(40)_{-40}^{+40} \times 10^{-4}$   &   $6.990(36)_{-42}^{+37} \times 10^{-4}$\\ 
                        Longitude of periastron ($^{\circ}$) & $\omega_{\rm p}$   &   $-7.06(16)_{-33}^{+33}$   &   $-7.06(20)_{-28}^{+29}$\\ 
                        Time of periastron passage ($\mathrm{MJD}$) & ${t_p}_{\rm I}$   &   $55917.1269(88)_{-15}^{+15} $   &   $55917.1269(86)_{-13}^{+13}$\\ 
                        Longitude of asc. node  ($^{\circ}$) & $\Omega_{\rm p}$   &   $145.9_{-1.5}^{+1.4} $   &   $148.0_{-1.5}^{+1.7} $\\ 
                        \hline
                        \multicolumn{4}{l}{\textit{Outer orbit $\rm O$ of $\rm b$ w.r.t. triple-system centre of mass $\rm t$}} \\
                        Semi-major axis (lt-s) & $a_{\rm b}$   &   $1.180(33)_{-29}^{+28} \times 10^{2}$   &   $1.180(27)_{-25}^{+27} \times 10^{2}$\\ 
                        Orbital inclination  ($^{\circ}$) & $i_{\rm O}$   &   $39.2(45)_{-11}^{+11}$   &   $39.2(48)_{-10}^{+10}$\\ 
                        Orbital eccentricity & $e_{\rm O}$   &   $0.03529117(20)_{-29}^{+29}$   &   $0.03529117(21)_{-28}^{+29} $\\ 
                        Longitude of periastron  ($^{\circ}$) & $\omega_{\rm b}$   &   $-5.73232(85)_{-27}^{+27}$   &   $-5.73232(56)_{-27}^{+27}$\\ 
                        Time of periastron passage ($\mathrm{MJD}$) & ${t_p}_{\rm O}$   &   $56224.98440(20)_{-19}^{+19} $   &   $56224.98440(42)_{-19}^{+19} $\\ 
                        \hline
                        Pulsar mass ($\Msol$) & $m_{\rm p}$   &   $1.43(82)_{-12}^{+12}$   &   $1.43(80)_{-11}^{+12}$\\ 
                        Inner-companion mass ($\Msol$) & $m_{\rm i}$   &   $0.197(58)_{-15}^{+15}$   &   $0.197(55)_{-14}^{+15}$\\ 
                        Outer-companion mass ($\Msol$) & $m_{\rm o}$    &   $0.410(18)_{-31}^{+31}$   &   $0.410(12)_{-29}^{+31}$
                \end{tabular}
        \end{footnotesize}
        \tablefoot{ Error bars apply to the digits between parenthesis and delimit the 68\% median confidence region, namely the interval excluding 16\% of the distribution above and below it. The central value quoted is the best-fit value. Upper-case indices $\rm I,O$ refer to the inner and outer binaries respectively. Lower-case indices $\rm p,i,o$ refer to the pulsar, inner white dwarf, and outer white dwarf, respectively; $\rm b$ refers to the inner-binary centre of mass. Data spans MJD $56492 - 59480$. Solar System ephemeris is DE430. Model-specific parameters are reported separately in Table \ref{tab:modelspecific}.}
\end{table*}

\end{appendix}

\end{document}